\documentclass[aps,preprintnumbers, supercriptaddress, amsmath, showpacs]{revtex4}
\usepackage{psfrag}
\usepackage{epsfig}
\usepackage{amsfonts} % para incluir fonts y
\usepackage{amssymb}  % simbolos de la AMS
\usepackage{graphicx} % standard LaTeX graphics tool
\usepackage{amsmath}  % for including eps-figure files
\usepackage{amsmath}  % paquete matemÃ¯Â¿Âœtico

\usepackage[latin1]{inputenc}
\usepackage{color}
\usepackage{suffix}
\usepackage{mathtools}

\newcommand{\bea}{\begin{eqnarray}}
\newcommand{\eea}{\end{eqnarray}}
\newcommand{\beas}{\begin{eqnarray*}}
\newcommand{\eeas}{\end{eqnarray*}}

\begin{document}

\title{Magnetic Dual Chiral Density Wave: A Candidate Quark Matter Phase for the Interior of Neutron Stars}

\author{E. J. Ferrer and V. de la Incera}
\affiliation{Dept. of Physics and Astronomy, Univ. of Texas Rio Grande Valley, 1201 West University Drive, Edinburg, Texas 78539, USA }

\begin{abstract}
In this review, we discuss the physical characteristics of the magnetic dual chiral density wave (MDCDW) phase of dense quark matter and argued why it is a promising candidate for the interior matter phase of neutron stars. The MDCDW condensate occurs in the presence of a magnetic field. It is a single-modulated chiral density wave characterized by two dynamically generated parameters: the fermion quasiparticle mass $m$ and the condensate spatial modulation $q$. The lowest Landau level quasiparticle modes in the MDCDW system are asymmetric about the zero energy, a fact that leads to the topological properties and anomalous electric transport exhibited by this phase. The topology makes the MDCDW phase robust against thermal phonon fluctuations, and as such, it does not display the Landau-Peierls instability, a stapled feature of single-modulated inhomogeneous chiral condensates in three dimensions. The topology is also reflected in the presence of the electromagnetic chiral anomaly in the effective action and in the formation of  hybridized propagating modes known as an axion-polaritons. Taking into account that one of the axion-polaritons of this quark phase is gapped, we argued how incident $\gamma$-ray photons can be converted into gapped axion-polaritons in the interior of a magnetar star in the MDCDW phase leading the star to collapse, a phenomenon that can serve to explain the so-called missing pulsar problem in the galactic center.
\end{abstract}

\maketitle

\section{Introduction}

 A fundamental question in nuclear physics/astrophysics nowadays is what is the state of matter that realizes in the interior of neutron stars (NS). Neutron stars are among the most dense objects in the universe. They are produced by the gravitational collapse of very massive stars that can have up to 30 solar masses or by binary NS merger events such as GW170817 \cite{Merger}. Their inner densities can reach values several times larger than the nuclear density $\rho_n=4\times10^{17} kg/m^3$. One possibility is that at those densities, baryons are so close that they  can be smashed together producing quark deconfinement.
Once the quark are liberated, there exists the  possibility to have NS exclusively formed by strange matter, the so called strange stars \cite{Weber}. The idea of a strange star was prompted by the Bodmer-Terazawa-Witten hypothesis \cite{Bodmer}-\cite{Witten} based on the idea that strange matter has a lower energy per baryon than ordinary nuclei even including $_{56}Fe$. Thus, the true ground state of the hadrons may be strange matter. Later on, the equilibrium composition and the equation of state (EoS) for strange matter were studied by other authors \cite{Farhi, Alcock, Xu-1, Xu-2, Paczynski, Weber}. Thus, a strange star will be formed by an absolutely stable phase consisting of roughly equal numbers of up, down, and strange quarks plus a smaller number of electrons (to guarantee charge neutrality). More recently, by using a phenomenological quark-meson model that includes the flavor-dependent feedback of the quark gas on the QCD vacuum, it was demonstrated in Ref. \cite{Holdom} that $u$-$d$ quark matter is in general more stable than strange quark matter, and it can be more stable than the ordinary nuclear matter when the baryon number is sufficiently large.

Considering effective models of Nambu-Jona-Lasinio (NJL) type with parameters matched to nuclear data, we can simulate the one-gluon exchange interaction of QCD, which contains a dominant attractive diquark channel. This attractive interaction gives rise to color superconductivity (CS) \cite{Reviews}-\cite{Hatsuda}.  NJL models have predicted that the most favored phase of CS at asymptotically high densities is the three-flavor color-flavor-locked (CFL) phase with a significant large gap. The existence of a large superconducting gap together with a repulsive vector interaction, which is always present in a dense medium \cite{Kitazawa2002}, can help to make the EoS stiffer enough to reach the high stellar masses measured for two compact objects, PSR J1614-2230 and PSR J0348+0432 with $M=1.97\pm 0.04M_{\odot}$ \cite{Demorest} and $M=2.01\pm 0.04M_{\odot}$ \cite{Antoniadis}, respectively, where $M_{\odot}$ is the solar mass. In a recent paper \cite{Odilon}, it was found that in addition to the previously cited agreement with respect to the stellar maximum mass, there is also a strong correlation between the predictions of the CFL model in a plausible range of parameters, even including the radiative effects of gluons \cite{Gluons}, and 
the mass/radius fits to NICER data for PSR J003+0451, as well as the tidal deformabilities of the GW170817 event.
 
Despite these encouraging results, the CFL phase fails to pass another important astrophysical test: the heat capacity lower limit obtained from temperature observations of accreting NS in quiescence. As found in \cite{Cv-NS}, the heat capacity of the NS core has a lower limit $\tilde{C}_V \geq10^{36}(T/10^8)$ erg/K. Thus, NS matter-phase candidates that do not satisfy this constraint should be ruled out.  Superfluid/superconducting phases where all the fermions are paired do not obey the constraint since they have a very small heat capacity proportional to $e^{-a/T}$ at small T, with $a$ a model-dependent function of the gap. Only superfluid/superconducting phases where not all the fermions are paired have the possibility to produce sufficiently heat capacity to satisfy the lower limit thanks to the contribution of non-paired fermions. These arguments were explicitly corroborated in \cite{Cv-NS, Pedro} for a pure CFL phase, showing that its heat capacity is strongly depleted not only because all the quarks form Cooper pairs, but also because the system does not have many electrons as its electrical neutrality is ensured by the almost equal numbers of $u$, $d$ and $s$ quarks alone. These results indicate that a pure CFL phase is not a suitable choice for the inner composition of compact stars. 

NS not only are the natural objects with the highest density in the universe, but they also exhibit the strongest magnetic fields, which become extremely large in the case of magnetars, with inner values that have been estimated to range from $10^{18}$ G for nuclear matter \cite{Nuclear-matter-field} to $10^{20}$ G for quark matter \cite{Quark-Matter-field}. The fact that strong magnetic fields populate the vast majority of the astrophysical compact objects and that they can significantly affect several properties of the star have served as motivation for many works focused on the study of the EoS of magnetized NS \cite{Quark-Matter-field}-\cite{Lattimer}. An important characteristics is that the EoS in a uniform magnetic field becomes anisotropic, with different pressures along the field and transverse to it \cite{Quark-Matter-field}-\cite{Aric}. A magnetic field has been shown to play an important role in CS \cite{Horvath}, \cite{CSB-1}-\cite{CSB-10}, as well as in inhomogeneous chiral phases \cite{Klimenko}-\cite{Bo}.

The presence of a magnetic field is relevant due to the activation of new channels of interaction and, occasionally, also due to the generation of additional condensates. For instance, in the quarkyonic phase of dense quark matter, a magnetic field is responsible for the appearance of a new chiral spiral between the pion and magnetic moment condensates, $\langle\bar{\psi}\gamma^5\psi\rangle$ and $\langle \bar{\psi}\gamma^1\gamma^2\psi\rangle$ respectively \cite{Quarkyonic}.  Similarly, additional condensates emerge in the homogeneous chiral phase \cite{AMM-NJL}, as well as in color superconductivity \cite{CSB-6}.

On the other hand, various QCD effective model studies, as well as QCD calculations in the large-Nc limit indicate that spatially inhomogeneous chiral phases, characterized by particle-holes pairs that carry total momentum, can be formed at relatively low temperatures and intermediate densities \cite{Rubakov}-\cite{Tatsumi-1}. Such inhomogeneous chiral phases emerge when the baryon density increases from low values, where the hadronic phase is favored, to densities a few times the nuclear saturation density. 

Interesting enough, approaching the low-temperature/ intermediate-density region from the other side, i.e. from the very high density region, also favors the formation of spatially inhomogeneous phases, only that in this case they are CS phases since their ground state contains quark-quark pairs \cite{Reviews, CS-MRP86}.  This phenomenon can be understood as follows. The CFL phase, favored at asymptotically large densities, is based on BCS quark-pairing. In this phase the quarks pair at the Fermi surface with equal and opposite momenta, so the phase is homogenous. However, with decreasing density, the combined effect of the strange quark mass, neutrality constraint and beta equilibrium, create a mismatch in the Fermi momenta of different flavors. The mismatch in turn imposes an extra energy cost on Cooper pair formation.  BCS-pairing can then dominate as long as the energy cost of forcing all species to have the same Fermi momentum is compensated by the win in pairing energy due to Cooper pair formation. The consequence of this competing effects is that eventually, as the density decreases, the CFL phase becomes the gapless CFL (gCFL) \cite{gCFL}, on which not all the Cooper pairs remain stable energetically anymore and as a consequence some of the quarks become gapless. More importantly,  the onset of gCFL produces chromomagnetic instabilities (CMI) \cite{CMI, gCFL-2}, meaning some of the gluons acquire imaginary Meissner masses, a sign that one is working in the wrong ground state. A viable solution, free of CMI, involves a momentum-dependent quark-quark condensate that spontaneously breaks translational invariance \cite{TBinCS, TBinCS-2, TBinCS-3, TBinCS-4} and hence form a spatially inhomogeneous CS phase. Most inhomogeneous CS phases are based on the idea of Larkin and Ovchinnikov (LO) \cite{LO} and Fulde and Ferrell (FF) \cite{FF}, originally applied to condensed matter. In the CS LOFF phases \cite{CSLOFF, CSLOFF-2, CSLOFF-3}, quarks of different flavors pair even though they have different Fermi momenta, because they form Cooper pairs with nonzero momentum. CS inhomogeneous phases with gluon vortices that break rotational symmetry \cite{gluonCS} have also been considered to remove the instability. 

Even though the above mentioned studies suggest that the inhomogeneous phases must be unavoidable at intermediate densities and low temperatures, the question of which phase is the most energetically favorable on each segment of the intermediate region still remains unanswered. Exploring it will required involved calculations due to the fact that the pairing energies between particle-particle, particle-antiparticle and particle-hole are comparable at these densities. 

In the present review,  we  shall  focus  our  attention  on  one  particular spatially inhomogeneous  phase,  a chiral phase known as the magnetic dual chiral density wave  (MDCDW) phase \cite{Klimenko}-\cite{Ferrer-NPB}. The MDCDW  ground  state is characterized by a chiral density wave made of scalar and pseudo-scalar condensate components, so the term "dual" in its name. This phase occurs in the presence of a magnetic field and exhibits a wealth of interesting topological properties. The MDCDW phase has profound differences with the so-called dual chiral density wave (DCDW) phase \cite{Tatsumi-1} where no external field is present, even though both are characterized by the same type of inhomogeneous chiral condensate. The magnetic field explicitly reduces the rotational and isospin symmetries that are present in the DCDW case, significantly enhances the window for inhomogeneity \cite{Klimenko}, and leads to topologically nontrivial transport properties \cite{Ferrer-PLB, Ferrer-NPB}. 

 An additional effect that makes the MDCDW phase a particularly viable candidate for the NS inner state of matter is that it is not washed out by thermal fluctuations at low temperatures.  This property is significant because even though single-modulated chiral condensates are energetically favored over their homogeneous counterpart at increasing densities, and favored even over higher-dimensional modulations in three dimensions, the long-order range in single-modulated condensates is always washed out by the thermal fluctuations of the Goldstone bosons at arbitrarily small temperatures. This occurs due to the existence of soft modes of the fluctuation spectrum in the direction normal to the modulation, a phenomenon known in the literature as Landau-Peierls instability \cite{Peirls, Landau}. In dense QCD models, the Landau-Peierls instability occurs in the periodic real kink crystal phase \cite{Hidaka}; in the DCDW phase \cite{Tatsumi}; and in the quarkyonic phase \cite{Pisarski}. The Landau-Peierls instability signals the lack of long-range correlations at any finite temperature and hence the lack of a true order parameter. Only a quasi long-range order remains in all these cases, a situation that resembles what happens in smectic liquid crystals \cite{smectic liquid crystal}. 

Thanks to the external magnetic field, the Landau-Peierls instability is absent in the MDCDW phase \cite{Incera}. The field produces two main effects. First, it acts as an external vector that explicitly breaks the rotational and isospin symmetries, allowing the formation of additional structures in the Ginzburg-Landau (GL) expansion of the MDCDW thermodynamic potential and reducing to one the number of Goldstone bosons in the spontaneously broken symmetry theory. Second, it induces a nontrivial topology in the system that manifests itself in the asymmetry of the lowest Landau level (LLL) modes and in the appearance of odd-in-q terms in the GL expansion. These two features in turn affect the low-energy theory of the thermal fluctuations, stiffening the dispersion relation in the direction normal to the modulation vector, thereby preventing the washout of the long-range order and hence removing the Landau-Peierls instability.

In this review, we will discuss the main properties of the MDCDW phase, including how the interaction of the MDCDW medium with an electromagnetic field modifies the propagation of electromagnetic waves there leading to interesting implications for astrophysics.

The review is organized as follows. In Section 2, we introduce the 2-flavor NJL model that serves as the basis for the MDCDW phase of dense quark matter in a magnetic field, outlining the derivations that lead to the emergence of a chiral anomaly term in the effective action of the system. In Section 3, we discuss the realization of axion electrodynamics in the MDCDW phase and the implications for electric transport. In Section 4, we demonstrate the lack of the Landau-Peierls instability in the MDCDW system and discuss the role played by the background magnetic field on this property. In Section 5, we go beyond the mean-field approximation to study the anomalous matter-light interaction that takes place in this inhomogeneous phase. We show how photons couple to the fluctuation of the axion field (proportional to the phonon fluctuation) to produce hybrid modes of propagation called axion polaritons. A possible consequence of the formation of these hybridized modes inside a quark star bombarded by $\gamma$-ray is then proposed in Section 6 to explain the so-called pulsar missing problem in the galactic center. Section 7 summarizes the main results and our concluding remarks.

%%%%%%%%%%%%%%%%%%%%%%%%%%%%%%%%%%%%%%%%%%
\section{The Magnetic Dual Chiral Density Wave Phase} \label{II}

To study the MDCDW phase, we start from a two-flavor NJL model of strongly interacting quarks at finite baryon density that includes the electromagnetic interaction and a background magnetic field
\begin{eqnarray} \label{L_NJL_QED}
\mathcal{L}=-\frac{1}{4}F_{\mu\nu}F^{\mu\nu}+\bar{\psi}[i\gamma^{\mu}(\partial_\mu+iQA_{\mu})+\gamma_0 \mu]\psi 
+G[(\bar{\psi}\psi)^2+(\bar{\psi}i\tau\gamma_5\psi)^2],
\end{eqnarray}
Here, $Q=\mathrm{diag} (e_u,e_d)=\mathrm{diag} (\frac{2}{3}e,-\frac{1}{3}e)$, $\psi^T=(u,d)$; $\mu$ is the quark chemical potential; and G is the four-fermion coupling. The electromagnetic potential $A^{\mu}$ is formed by the background $\bar{A}^{\mu}=(0,0,Bx,0)$, which corresponds to a constant and uniform magnetic field $\mathbf{B}$ pointing in the z-direction, with $x^{\mu}=(t,x,y,z)$, and a fluctuation field $\tilde{A}$. Because of the electromagnetic coupling, the flavor symmetry in this model is $U(1)_L\times U(1)_R $. In addition, the background magnetic field explicitly breaks the rotational symmetry that exists in its absence, so that the spatial symmetry of (\ref {L_NJL_QED}) is $SO(2)\times R^3$.

It has been shown that at finite baryon density, the two condensates
\begin{equation}\label{DCDW-cond}
\langle\bar{\psi}\psi\rangle= \Delta \cos q_{\mu}x^{\mu}, \qquad \langle\bar{\psi}i\tau_3 \gamma_5\psi\rangle= \Delta \sin{q_{\mu}x^{\mu}} ,
\end{equation}
gets expectation values different from zero forming a dual chiral density wave condensate, with its modulation vector favored along the field direction $q^{\mu}=(0,0,0,q)$ \cite{Klimenko, Tatsumi-2}. Notice that this means that the modulation is $q$ for the u-quarks and $-q$ for the d-quarks.

Expanding the Lagrangian (\ref{L_NJL_QED}) about  this inhomogenous condensate, bozonizing the four-fermion interaction via the Hubbard-Stratonovich approach, and taking the local chiral transformations
 \begin{equation}\label{chiralT}
\psi \to e^{i\tau_3\gamma_5\theta}\psi, \quad \bar{\psi} \to \bar{\psi}e^{i\tau_3\gamma_5 \theta}
\end{equation}
 with $\theta=qz/2$, we arrive at the mean-field Lagrangian 
\begin{equation}\label{U_1-MF_L}
\mathcal{L}_{MF}=\bar{\psi}[i\gamma^{\mu}(\partial_\mu+iQA_{\mu}+i\tau_3\gamma_5\partial_{\mu}\theta) +\gamma_0\mu-m]\psi-\frac{m^2}{4G}-\frac{1}{4}F_{\mu\nu}F^{\mu\nu}
\end{equation}
where $m=-2G\Delta$, thus the quasiparticle mass is proportional to the condensate magnitude.
 
The energy spectrum of the theory (\ref{U_1-MF_L}) separates in two sets:

(1)  lowest Landau level (LLL) ($l=0$) 
\begin{equation}\label{LLLspectrum}
E^{0}=\epsilon\sqrt{m^2+k_3^2}+q/2,  \quad \epsilon=\pm,
\end{equation}

(2) higher Landau levels (HLL) ($l\neq0$) 
\begin{equation}\label{HighLspectrum-1}
E^l= \epsilon\sqrt{(\xi\sqrt{m^2+k_3^2}+q/2)^2+2|e_fB|l}, \quad \epsilon=\pm, \xi=\pm, l=1,2,3,...
\end{equation}

The HLL spectrum has four branches, with $\xi=\pm$ indicating spin projections and $\epsilon=\pm$ the energy sign. In contrast, the LLL has only two branches because only one spin projection contributes to the LLL modes. Here, $\epsilon$ loses the energy sign interpretation as long as $q \ne 0$ \cite{Klimenko}.  An important feature of this spectrum is that the LLL energies are not symmetric about the zero-energy level. This asymmetry in the LLL spectrum gives rise to nontrivial topological effects that will be pointed out below.

It is important to note that the fermion measure in the path integral is not invariant under the local chiral transformation (\ref{chiralT}) and hence it produces a contribution to the  action through the transformation's Jacobian 
$J(\theta(x))=(\textrm{Det}U_A)^{-2}$ 
\begin{equation}\label{MeasureNI}
\mathcal{D}\bar{\psi}(x)\mathcal{D}\psi(x) \to (\textrm{Det}U_A)^{-2} \mathcal{D}\bar{\psi}(x)\mathcal{D}\psi(x),
\end{equation} 
with  $U_A=e^{i\tau_3\gamma_5\theta}$. However, $J(\theta(x))$ is ill-defined and needs to be regularized.  This can be done using the Fujikawa method \cite{Fujikawa}, so that the measure contribution to the mean-field action turns out to be an axion term given by the electromagnetic chiral anomaly $\frac{\kappa}{4} \theta(x) F_{\mu\nu}\tilde{F}^{\mu \nu}$ \cite{Ferrer-PLB}- \cite{Ferrer-NPB}. Then,
 \begin{eqnarray}\label{Action-Effective}
S_{eff}&=&\int d^4x \{\bar{\psi}[i\gamma^{\mu}(\partial_\mu+iQA_{\mu}+i\tau_3\gamma_5\partial_{\mu}\theta) +\gamma_0\mu-m]\psi
-\frac{m^2}{4G}\nonumber
\\
&+&\frac{\kappa}{4} \theta(x) F_{\mu\nu}\tilde{F}^{\mu \nu}-\frac{1}{4}F_{\mu\nu}F^{\mu\nu}\},
\end{eqnarray}
The coupling between the background axion field $\theta(x)$ and the electromagnetic tensor is given by $\frac{\kappa}{4}=\frac{3(e_u^2-e_d^2)}{8\pi^2}=\frac{e^2}{8\pi^2}=\frac{\alpha}{2\pi}$. It contains the contribution of all the quark flavors and colors. 

The one-loop thermodynamic potential of the mean-field theory was found in Refs. \cite{Klimenko, Ferrer-NPB} to be 
\begin{equation}\label{Thermodynamic-Pot}
\Omega=\Omega_{vac}(B)+\Omega_{anom}(B,\mu)+\Omega_\mu(B,\mu)+\Omega_T(B, \mu, T)+\frac{m^2}{4G}, 
\end{equation}
Where $\Omega_{vac}$ is the vacuum contribution; $\Omega_{anom}$ is the anomalous contribution, extracted from the LLL part of the medium term after proper regularization  \cite{Klimenko}; $\Omega_\mu$ is the zero-temperature medium contribution and $\Omega_T$  the thermal contribution. For a single quark flavor $f$ they are \cite{Ferrer-NPB}
\begin{eqnarray}\label{TP-Contributions}
\Omega^f_{vac}&=&\frac{1}{4\sqrt{\pi}} \frac{N_c|e_fB|}{(2\pi)^2}\int_{-\infty}^\infty dk\sum_{l\xi\epsilon} \int^{\infty}_{1/\Lambda^2} \frac{ds}{s^{3/2}}e^{-s(E)^2}
\\
\Omega^f_{anom} &=&- \frac{N_c|e_fB|}{(2\pi)^2} q\mu \label{Omega-anom}
\\
\Omega^f_\mu&=&-\frac{1}{2} \frac{N_c|e_fB|}{(2\pi)^2}\int_{-\infty}^\infty dk\sum_{\xi,l>0}2\lbrack(\mu-E)\Theta(\mu-E)\rbrack\vert_{\epsilon=+}\nonumber
\\
&\quad&+\Omega^{fLLL}_\mu
\\
\Omega^f_T&=& -\frac{N_c|e_fB|}{(2\pi)^2\beta}\int_{-\infty}^{\infty} dk\sum_{l\xi\epsilon}\ln \left (1+e^{-\beta (|E-\mu|} \right )
\end{eqnarray} 
with $E$ the energy modes (\ref{LLLspectrum}) and (\ref{HighLspectrum-1}), and the LLL zero-temperature medium contribution given by
\begin{eqnarray}\label{Medium-Contribution-2}
\Omega^{fLLL}_\mu&=&-\frac{1}{2} \frac{N_c|e_fB|}{(2\pi)^2}\int_{-\infty}^{\infty} dk\sum_\epsilon(|E^0-\mu|-|E^0|)_{reg}
\\
&=&-\frac{N_c|e_fB|}{(2\pi)^2} \Bigg\{\Bigg[ Q(\mu) + m^2\ln \bigg(m/R(\mu)\bigg)\Bigg ] \Theta(q/2-\mu-m)\Theta(q/2-m)\nonumber
\\
&-&\Bigg[ Q(0) + m^2\ln \bigg(m/R(0)\bigg)\Bigg ] \Theta(q/2-m)\nonumber
\\
&+&\Bigg[ Q(\mu) + m^2\ln \bigg(m/R(\mu)\bigg)\Bigg ]  \Theta(\mu-q/2-m)\nonumber
\\
&-&\Bigg[ Q(0) + m^2\ln \bigg(m/R(0)\bigg)\Bigg ]  \Theta(\mu-q/2-m)\Theta(-q/2-m)\Bigg\},\nonumber
\end{eqnarray}
 Here, we introduced the notation
\begin{eqnarray}\label{Parameter definitions}
Q(\mu)&=&|q/2-\mu|\sqrt{(q/2-\mu)^2-m^2}, \quad \quad Q(0)=|q/2|\sqrt{(q/2)^2-m^2}\nonumber
\\
R(\mu)&=&|q/2-\mu|+\sqrt{(q/2-\mu)^2-m^2}, \quad R(0)=|q/2|+\sqrt{(q/2)^2-m^2}\nonumber
\end{eqnarray}

Notice that the anomalous term $\Omega^f_{anom}$ favors a nonzero modulation $q$ since it decreases the free-energy of the system. Such a term is a direct consequence of the asymmetry of the LLL spectrum and hence has a topological origin. 

The minimum solutions for $m$ and $q$ in terms of the chemical potential and the external magnetic field can be found by numerically solving the gap equations \cite{Klimenko} and \cite{Bo}
\begin{equation}\label{Gap-Eq}
\frac{\partial\Omega}{\partial m}=0, \quad  \quad  \frac{\partial \Omega}{\partial q}=0.
\end{equation}

 %%%%%%%%%%%%%%%%%%%%%%%%%%%%%%%%%%%%
\begin{figure}
\begin{center}
\begin{tabular}{ccc}
  \includegraphics[width=7.5cm]{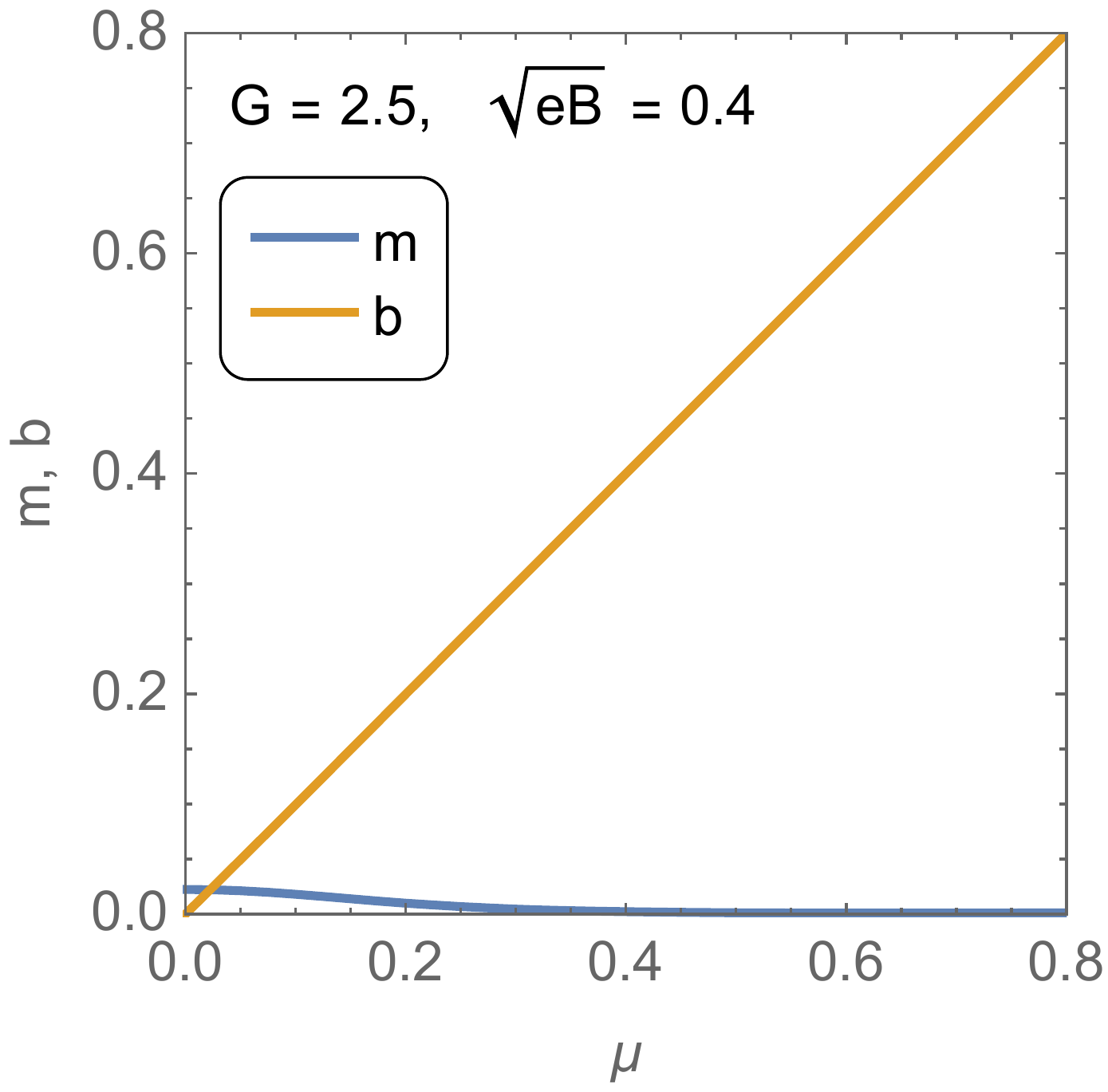} & \includegraphics[width=7cm]{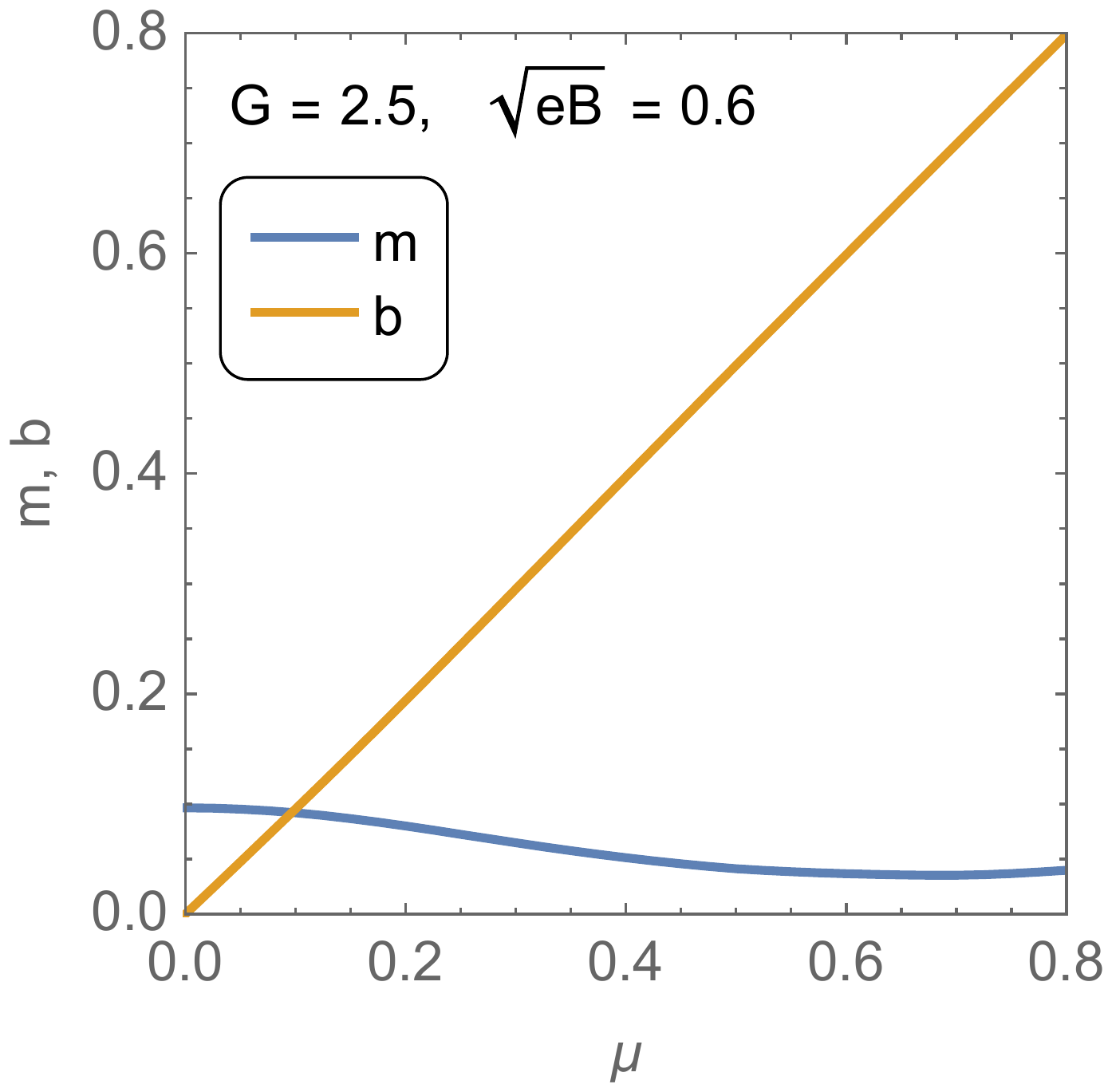}\\
(a) & (b)  \\
  \\
  \end{tabular}
    \end{center}
    \caption{Solutions of the MDCDW gap equations versus the quark chemical potential at subcritical coupling ($G= 2.5$) and magnetic fields ($\sqrt{eB}=0.4, 0.6$).}
     \label{Fig-1}
\end{figure}
%%%%%%%%%%%%%%%%%%%%%%%%%%%%%%%%%%%%

 %%%%%%%%%%%%%%%%%%%%%%%%%%%%%%%%%%%%
\begin{figure}
\begin{center}
\begin{tabular}{ccc}
  \includegraphics[width=7cm]{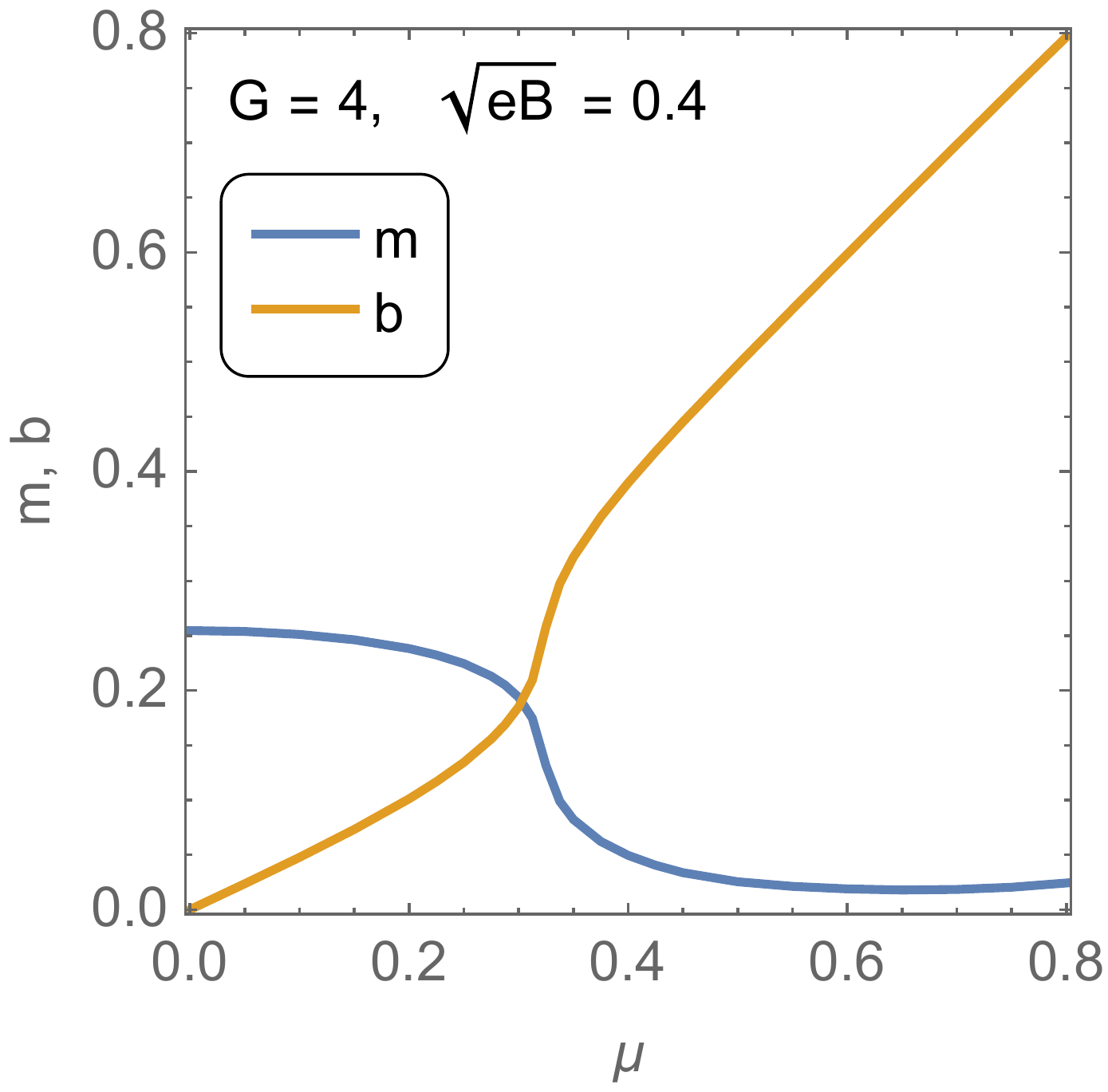} & \includegraphics[width=7cm]{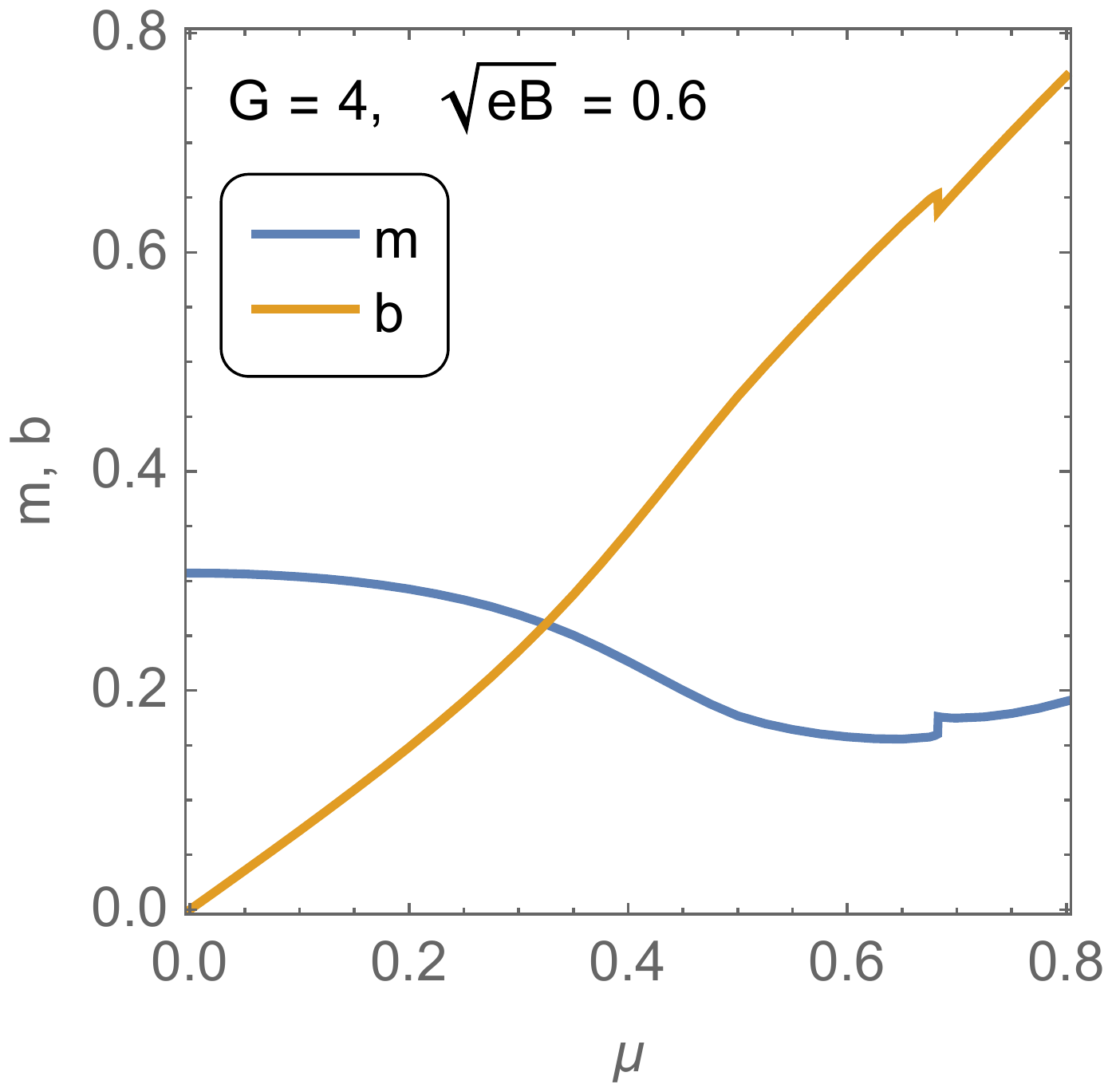}\\
(a) & (b)  \\
  \\
  \end{tabular}
    \end{center}
    \caption{Solutions of the MDCDW gap equations versus the quark chemical potential at supercritical coupling ($G= 4$) and magnetic fields ($\sqrt{eB}=0.4, 0.6$).}
     \label{Fig-2}
\end{figure}
%%%%%%%%%%%%%%%%%%%%%%%%%%%%%%%%%%%%

 Fig.1 shows the resulting $m$ and $b=q/2$ vs $\mu$ at undercritical coupling $G=2.5$ for two strong magnetic strengths. Thanks to the magnetic field, the MDCDW solution exists even in the undercritical regime. Notice that the condensate magnitude is quite sensitive to the change in the field strength, while its modulation is not. For $\sqrt{eB}=0.4$, $m$ is at least one order of magnitude smaller than for $\sqrt{eB}=0.6$ for the entire range of $\mu$ considered. Fig. 2 shows the solutions in the supercritical case. In this case, the effect of increasing the magnetic field is also noticeable in $m$ but still not significant in $b$, except that increasing the field tends to smooth out the behavior of the dynamical parameters in the region before and after they cross each other.  Comparing the two set of curves, it becomes apparent that at a given magnetic field, larger coupling leads to larger condensate magnitude, but not larger modulation. All the quantities in the figures are normalized with respect to the proper-time regularization parameter $\Lambda=636.790$ MeV, thus are dimensionless.
 
%%%%%%%%%%%%%%%%%%%%%%%%%%%%%%%%%%%%%%%%%%
\section{Electromagnetism in the MDCDW Phase}

To obtain the electromagnetic effective action  $\Gamma(A)$ in the MDCDW phase we start from the formula
\begin{equation}
\Gamma=-i\log Z,
\end{equation}
where the partition function $Z$ is
\begin{equation}\label{Z-theta}
Z=e^{i\Gamma}= \int \mathcal{D}\bar{\psi}(x)\mathcal{D}\psi(x) e^{iS_{eff}}
\end{equation}
with $S_{eff}$ given in (\ref{Action-Effective}).

Integrating in the fermion fields, performing the Matsubara sum, taking the zero-temperature limit, and  expanding $\Gamma$ in powers of the fluctuation field $\tilde{A}$ we obtain
\begin{eqnarray} \label{EA-2}
\Gamma(A)&=&-V\Omega+\int d^4x \left[-\frac{1}{4}F_{\mu\nu}F^{\mu\nu}+\frac{\kappa}{4} \theta(x) F_{\mu\nu}\tilde{F}^{\mu \nu}\right]
\\
&+ &\sum_{i=1}^{\infty}\int dx_{1}...dx_{i}\Pi^{\mu_1,\mu_2,...\mu_i}(x_1,x_2,...x_i)\tilde{A}_{\mu_1}(x_1)...\tilde{A}_{\mu_i}(x_i),\nonumber
\end{eqnarray}
with V the four-volume, $\Omega$ the mean-field thermodynamic potential in the one-loop approximation (\ref{Thermodynamic-Pot}), and $\Pi^{\mu_1,\mu_2,...\mu_i}$ the i-vertex tensors corresponding to the one-loop polarization operators with internal lines of fermion Green functions in the MDCDW phase and $i$ external lines of photons. 

We are interested in the linear response of the MCDCW phase to a small electromagnetic probe $\tilde{A}$. For consistency of the approximation, we can neglect all the radiative corrections of order higher than $\alpha$, as $\alpha$ is the order of the axion term in (\ref{EA-2}).  These conditions imply that we shall cut the series at $i=1$, which can be shown to provide the medium corrections to the Maxwell equations that are linear in the electromagnetic field and in $\alpha$.

Hence, the electromagnetic effective action becomes  
\begin{eqnarray} \label{EA}
\Gamma(A)&=&-V\Omega+\int d^4x \left[-\frac{1}{4}F_{\mu\nu}F^{\mu\nu}-\kappa\int d^4x  \epsilon^{\mu \alpha \nu \beta}A_\alpha \partial_\nu A_\beta\partial_{\mu} \theta \right]
\nonumber
\\
&-&\int d^4x \tilde{A}_\mu(x) J^\mu(x) ,
\end{eqnarray}
where we integrated by parts the third term in the r.h.s. of (\ref{EA-2}).  The four-current $J^\mu(x)=(J^0,\mathbf{J})$ represents the contribution of the ordinary (non-anomalous) electric four-current, obtained from the one-loop tadpole diagrams. 

The Euler-Lagrange equations derived from this effective action turn out to be the equations of axion-electrodynamics
\begin{eqnarray} \label{axionMaxwell}
&\mathbf{\nabla} \cdot \mathbf{E}=J^0+\frac{e^2}{4\pi^2}qB, \label{1-1}
 \\
&\nabla \times \mathbf{B}-\partial \mathbf{E}/\partial t=\mathbf{J}-\frac{e^2}{4\pi^2} \mathbf{q}\times \mathbf{E},  \label{2-2}
\\
&\mathbf{\nabla} \cdot \mathbf{B}=0, \quad \nabla \times \mathbf{E}+\partial \mathbf{B}/\partial t=0 \label{3-3},
\end{eqnarray} 
where we already used $\theta=\frac{qz}{2}$ \cite{Ferrer-NPB}. Hence, electromagnetism in the MDCDW phase is described by a particular case of the axion-electrodynamic equations proposed many years ago for a general axion field $\theta$ \cite{Wilczek} . 

The $q$-dependent terms in (\ref{1-1})-(\ref{2-2}) are directly connected to the chiral anomaly and thus give rise to an anomalous electric charge density, 
\begin{equation}\label{anomalous-Charge}
J_{anom}^0=\frac{e^2}{4\pi^2}qB,
\end{equation}
and to an anomalous Hall current density, 
\begin{equation}\label{anomalous-Current}
\mathbf{J}_{anom}=-\frac{e^2}{4\pi^2} \mathbf{q}\times \mathbf{E}.
\end{equation}
The anomalous electric charge density (\ref{anomalous-Charge}) can be also found multiplying the flavor electric charge $e_f$ by the anomalous quark number density of that flavor, obtained as the derivative of $\Omega^f_{anom} =- \frac{N_c|e_fB|}{(2\pi)^2} q\mu$ with respect to $\mu$, and then summing in flavor \cite{Ferrer-NPB}. As it should be, the anomalous Hall current $\mathbf{J}_{anom}$ is perpendicular to the background magnetic field and the probe electric field, since $\textbf{q}$ is aligned with $\mathbf{B}$. 
 
In (\ref{axionMaxwell}), $J^0$ and $\mathbf{J}$ denote ordinary charge and current densities respectively, which are calculated through radiative corrections. The contribution of the LLL to the ordinary charge density can be found from the tadpole diagram \cite{Ferrer-PLB, Ferrer-NPB} and is
\begin{eqnarray}\label{regular-charge}
J_{LLL}^0&=&\sum_f J^0_{LLL}(\mathrm{sgn}\left(e_f\right))
\\
&=&\frac{e^2B}{2\pi^2}\sqrt{(\mu-q/2)^2-m^2}[\Theta(\mu-q/2-m)-\Theta(q/2-\mu-m)],\nonumber
\end{eqnarray} 

Since the  LLL ordinary charge density is linear in the magnetic field, one can use the Str\v{e}da formula \cite{Streda, Ferrer-PLB, Ferrer-NPB} \begin{equation}\label{HC-formula}
\sigma_{xy}=\frac{\partial J^0}{\partial B}
\end{equation} 
to show that the LLL  contribution to the ordinary Hall conductivity is given by

\begin{equation}\label{ordHC}
\sigma^{ord}_{xy}= \frac{\partial J_{LLL}^0}{\partial B} =\frac{e^2}{2\pi^2}\sqrt{(\mu-q/2)^2-m^2}[\Theta(\mu-q/2-m)-\Theta(q/2-\mu-m)],
\end{equation}
which in turn leads to the LLL ordinary Hall current 
\begin{equation}\label{ordHallcurrent}
\mathbf{J}_{LLL}^{ord}=(\sigma^{ord}_{xy}E_y,-\sigma^{ord}_{xy}E_x,0).
 \end{equation}
 
Likewise, the anomalous Hall conductivity can be found either from the anomalous charge (\ref{anomalous-Charge}),
\begin{equation}\label{anomHC}
\sigma^{anom}_{xy}= \frac{\partial J_{anom}^0}{\partial B} =\frac{e^2}{4\pi^2}q,
\end{equation} 
or directly from the anomalous Hall current (\ref{anomalous-Current}). As $J_{anom}^0$ is due to the LLL, so is $\sigma^{anom}_{xy}$, thereby underlining once again the LLL origin of  $\mathbf{J}_{anom}$.

The anomalous Hall conductivity has a topological origin since it is a direct consequence of the chiral anomaly. That means that it has a universal character and as such it is robust against dissipative effects. This is quite analogous to what occurs in Weyl semimetals \cite{WeylSem}, where an anomalous Hall conductivity very similar to (\ref{anomHC}) is also connected to the chiral anomaly.  The only difference is that there the modulation $q$ is replaced by the separation in momentum between the two Weyl nodes.  A more important difference is that in Weyl semimetals a gap term that explicitly breaks chiral symmetry may exist, in contrast to the MDCDW where the theory is initially massless and the mass $m$ is dynamically generated by the spontaneous breaking of chiral symmetry induced by the inhomogeneous condensate. Even when there is an initial gap, there are still gapless Weyl points, as long as the separation between them is larger than the gap. Hence, even in this case, the anomalous Hall current of the Weyl semimetal is given by the same expression and remains robust against impurity scattering potentials, electron-electron interactions, or other similar dissipative effects \cite{WeylSem}. 

Something worth to notice is that the LLL contribution to the ordinary charge $J_{LLL}^0$ and Hall current $\mathbf{J}_{LLL}^{ord}$,  do not cancel out their corresponding anomalous counterparts (\ref{anomalous-Charge}) and (\ref{anomalous-Current}) \cite{Ferrer-PLB, Ferrer-NPB}, in sharp contrast to what occurs in the chiral magnetic effect in equilibrium where the anomalous and ordinary currents completely cancel out \cite{CME}. Nevertheless, in the limit when the order parameter $m$ becomes very small, $m\ll \frac{q}{2} < \mu$, one can expand the square root in the LLL ordinary part of the electric charge (\ref{regular-charge}) to see that the anomalous contribution gets effectively canceled out by one of the terms from the expansion of the ordinary part, leaving only terms that explicitly depend on $\mu$ and hence non-topological. The same type of cancellation happens between the anomalous Hall conductivity and a term coming from expanding the LLL ordinary Hall conductivity at very small $m$. This occurs near the phase transition line that separates the MDCDW phase from the chirally restored phase, where it is physically expected that the topology (or lack of it) at the two sides of the transition line should match. Therefore, the topological properties of the MDCDW phase become practically inoperative near the phase transition.

We point out that in Refs. \cite{Tatsumi-2},\cite{Yoshiike} a different method was employed to obtain the anomalous contributions to the fermion number, electric charge, and Hall current of the MDCDW phase. That method was based on the regularized Atiyah-Patodi-Singer index $\eta_H=\lim_{s\to 0}\sum_{l} \mathrm{sgn} (\lambda_{l}) \lambda_{l}^{-s}$ using an approach discussed in \cite{Niemi-Semenoff}. As found in \cite{Tatsumi-2}, the index gives different results for the anomalous fermion number depending on whether $m> q/2$ or viceversa. Since $m>q/2$ occurs at low chemical potentials, where no Fermi surface is generated, while $m<q/2$ occurs at chemical potentials large enough for a Fermi surface to exist, these results seem to indicate that there is an additional contribution to the anomalous fermion number in the region of high chemical potentials. Specifically, when $m>q/2$, Ref. \cite{Tatsumi-2} found that $\eta_H=-\frac{|eB|q}{2\pi^2}$, while when $m<q/2$ it was $\eta_H=\frac{|eB|}{2\pi} [-\frac{q}{\pi}+\frac{\sqrt{q^2-4m^2}}{\pi}]$. Based on these results, Ref. \cite{Yoshiike} claimed that a similar additional contribution entered in the anomalous Hall conductivity. Since such a term does not appear when one extracts the anomalous fermion number contribution using an energy cutoff regularization, as done in \cite{Klimenko} and \cite{Ferrer-PLB, Ferrer-NPB}, the authors of \cite{Yoshiike} concluded that the energy cutoff method is not good to extract the complete anomalous parts of physical parameters as the fermion number, Hall conductivity, electric charge, etc. What the authors of \cite{Yoshiike} failed to realize is that the additional term they found using the index approach and that they interpreted as the "anomalous" Hall conductivity for the region of $m<q/2$ (Eq. (18) in \cite{Yoshiike}) not only is not anomalous, but it is actually eliminated by an equal and opposite contribution coming from the ordinary part of the Hall conductivity (Eq. (20) in \cite{Yoshiike}). So the actual anomalous Hall conductivity is $\sigma^{anom}_{xy}=\frac{e^2}{4\pi^2}q$. Not just it is the same in all the regions, but it can be correctly extracted from the anomalous charge derived using the energy cutoff regularization approach employed in \cite{Klimenko}, or from the chiral anomaly obtained using the Fujikawa approach as done in \cite{Ferrer-PLB, Ferrer-NPB}.

As the above discussion illustrates, in the region of large chemical potentials the regularized Atiyah-Patodi-Singer index may contain, besides the genuinely anomalous part, some spurious non-anomalous contributions that cancel out with others coming from the ordinary part of the fermion number. To extract the correct anomalous contribution using the regularized index, one has to be particularly careful when using Niemi's approach for theories with finite chemical potential \cite{Niemi}. Indeed, one first has to add the index and the ordinary (Fermi surface) contributions to the fermion number, since only after that it is possible to cancel any spurious terms and then correctly separate the anomalous from the non-anomalous contributions in the fermion number and similarly in other quantities like the Hall conductivity. On the other hand, the advantage of finding the anomalous fermion number and electric charge with the energy cutoff approach or the anomalous charge and current from the chiral anomaly is that these approaches manage to extract the actual anomalous contribution without producing spurious terms.

 Another interesting property of the MDCDW medium becomes apparent by rewriting equations (\ref{1-1}) and (\ref{2-2}) in terms of the $D$ and $H$ fields 
\begin{equation}\label{eqs}
\mathbf{\nabla} \cdot \mathbf{D}=J^0,\quad \nabla \times \mathbf{H}-\frac{\partial \mathbf{D}}{\partial t}=\mathbf{J} 
\end{equation} 
which shows that in this model the field $\mathbf{D}$ and $\mathbf{H}$ are 
\begin{equation}\label{D-Hfields}
\mathbf{D}=\mathbf{E}-\kappa\theta \mathbf{B}, \quad \mathbf{H}=\mathbf{B}+\kappa\theta \mathbf{E}
\end{equation}
with $\kappa$ and $\theta$ defined in Section 2.
Equations (\ref{D-Hfields}) show that a magnetic field induces an electric polarization $\mathbf{P}=-\kappa\theta\mathbf{B}$ and an electric field induces a magnetization $ \mathbf{M}=-\kappa\theta\mathbf{E}$, a phenomenon known as magnetoelectricity. The linear magnetoelectricity of the MDCDW medium is a direct consequence of the chiral anomaly. It reflects the fact that the ground state of the MDCDW medium breaks Parity and Time inversion symmetries.
The magnetoelectricity in the MDCDW phase is different from the one found in the magnetic-CFL  phase of CS, where parity was not broken and the effect was a consequence of an anisotropic electric susceptibility \cite{CSB-7}, thus not linear. It also follows from (\ref{eqs}) that the anomalous Hall current is given by a medium-induced, magnetic current density $\nabla \times \mathbf{M}$, due to the space-dependent anomalous magnetization coming from the axion term. 

%We should mention that the same expression of the anomalous Hall conductivity has been found in Weyl semimetals \cite{WSM-1, WSM-2}, where the role of the modulation parameter $q$ is played by the separation in momentum of the Weyl nodes. This analogy is pointing out that in spite of the difference between these two systems that are realized in very different energy ranges, they are strongly related by their topological properties.

The above results might have some connotation for astrophysics. If the MDCDW phase is realized in the interior of NS, any electric field in the medium whether due to local separation of charges or any other possible reason, could trigger dissipationless Hall currents in the plane perpendicular to the magnetic field.  This current in turn could have a back effect on the magnetic field. Currents of this type could serve to resolve the issue about the stability of the magnetic field strength in magnetars  \cite{B-Stability-1, B-Stability-2}. 

%%%%%%%%%%%%%%%%%%%%%%%%%%%%%%%%%%%%%%%%%%
\section{Condensate Stability at Finite Temperature}

Let us discuss now how the finite temperature can affect the inhomogeneous condensate. 
As mentioned in the Introduction, single-modulated phases in three spatial dimensions exhibit the Landau-Peierls instability \cite{Peirls, Landau}. The Landau-Peierl instability is characterized by the fact that at nonzero temperatures, thermal fluctuations of the Nambu-Goldstone bosons, whose dispersions are anisotropic and soft in the direction normal to the modulation vector, wash out the long-range order at any finite temperature, signaling the lack of a true order parameter. Some inhomogeneity may remains, however, due to algebraically decaying long-range correlations of the order parameter, forming a phase with a quasilong range order similar to smectic liquid crystals \cite{smectic liquid crystal}. Depending on the size of the system, this much smoother inhomogeneity may or may not be relevant for the observables. 

Nevertheless, the presence of a magnetic field changes the properties of the low-energy theory in such a way that it completely removes the Landau-Peierls instability \cite{Incera}. To show that, we start from the low-energy theory of the MDCDW phase, described by a generalized GL expansion of the thermodynamic potential in powers of the order parameter and its derivatives. In the context of NS astrophysics, the region of interest is that of intermediate chemical potentials and low temperatures. Henceforth, we focus our investigation on that region, and work near the phase transition to the chirally restored phase. 

The validity of the GL expansion in this region is justified by the fact that the order parameters satisfy $m/\mu\ll1$  and $q/2\mu<1$ \cite{Klimenko}. One can readily show \cite{footnote}, following an approach similar to the one used in \cite{PRD97-036009} for the DCDW case, that the power series in $q$ effectively becomes an expansion in powers of $q/2\mu$, hence corroborating the consistency of the expansion and the truncation used. The GL expansion of the MDCDW phase near the critical point (CP), that is, in the region of large temperatures and low chemical potentials, was explored in \cite{Tatsumi-2}.

The GL expansion in our case should reflect the invariance with respect to the symmetries of the theory in the presence of the external magnetic field. In the MDCDW system, the order parameter is characterized by the scalar and pseudoscalar fields $\sigma= -2G\bar{\psi}\psi$ and $\pi=-2G\bar{\psi} i\gamma^5\tau_3\psi$, respectively. Under a global chiral transformation $e^{i\gamma_5 \tau_3\theta/2}$ of the fermion fields, they transform as $\sigma \to \sigma \cos \theta+\pi \sin \theta$ and $\pi \to \pi \cos \theta- \sigma \sin \theta$, reflecting the isomorphism between the chiral group  $U_A(1)$ and the $SO(2)$ of internal rotations acting on the two-dimensional vector $\phi=(\sigma, \pi)$. In a similar way, one can see that the $U_V(1)$ transformation of the fermions reduces to the trivial group acting on the vector $\phi$. 

Therefore, the GL expansion, in the SO(2) representation, can be written as
\begin{eqnarray}\label{GL-Free Energy-phi}
\mathcal{F}&=&a_{2,0}\phi^T\phi+\frac{b_{3,1}}{2}\left [\phi^T\hat{B}\cdot \widetilde{\nabla} \phi+\hat{B}\cdot (\widetilde{\nabla} \phi)^T \phi\right ]+a_{4,0} (\phi^T\phi)^2\nonumber
\\
&+&a^{(0)}_{4,2}(\widetilde{\nabla}\phi)^T\cdot \widetilde{\nabla}\phi+a^{(1)}_{4,2} \hat{B}\cdot (\widetilde{\nabla} \phi)^T \hat{B}\cdot \widetilde{\nabla} \phi \nonumber
\\
&+&\frac{b_{5,1}}{2}(\phi^T\phi)\left [\phi^T\hat{B}\cdot \widetilde{\nabla} \phi+\hat{B}\cdot (\widetilde{\nabla} \phi)^T \phi \right ]+ \frac{b_{5,3}}{2}\left [( \widetilde{\nabla}^2 \phi)^T\hat{B}\cdot \widetilde{\nabla}\phi+\hat{B}\cdot (\widetilde{\nabla}\phi)^T \widetilde{\nabla}^2 \phi \right ]+a_{6,0}(\phi^T\phi)^3 \nonumber
\\
&+&a^{(0)}_{6,2}(\phi^T\phi)(\widetilde{\nabla}\phi)^T\cdot \widetilde{\nabla}\phi+a^{(1)}_{6,2}(\phi^T\phi) [\hat{B}\cdot (\widetilde{\nabla} \phi)^T \hat{B}\cdot \widetilde{\nabla} \phi]+a_{6,4}(\widetilde{\nabla}^2 \phi)^T(\widetilde{\nabla}^2 \phi)+... ,
\end{eqnarray}
where we introduced the additional structural terms that are consistent with the symmetry of the theory in a magnetic field. The notation $\hat{B}=\mathbf{B}/|\mathbf{B}|$ for the normalized vector  in the direction of the magnetic field was used
and the gradient operator $-i\nabla$ in the SO(2) representation was introduced as
\begin{equation}
\widetilde{\nabla}= \left(\begin{array}{ccc}0 & 1 \\-1 & 0\end{array}\right)\nabla \qquad .
\end{equation}

The coefficients $a$ and $b$ are functions of $T, \mu$ and $B$. They can be derived from the MDCDW thermodynamic potential (\ref{Thermodynamic-Pot}) found in \cite{Klimenko, Ferrer-NPB}, although their explicit expressions are not relevant for the present study. The first subindex in the coefficients $a$ and $b$ indicates the power of the order parameter plus its derivatives in that term, the second index denotes the power of the derivatives alone.

We can now get advantage of the isomorphism between SO(2) and $U_A(1)$ to represent the order parameter as a complex function $M(x)=\sigma(x)+ i \pi(x)$. In terms of $M(x)$ the GL expansion of the free energy (\ref{GL-Free Energy-phi}) takes the form
\begin{eqnarray}\label{GL-Free Energy}
\mathcal{F}&=&a_{2,0}|M|^2-i\frac{b_{3,1}}{2}\left [M^*(\hat{B}\cdot \nabla M)-(\hat{B}\cdot \nabla M^*)M\right ]+a_{4,0} |M|^4+a^{(0)}_{4,2}|\nabla M|^2
\nonumber
\\
&+&a^{(1)}_{4,2} (\hat{B}\cdot \nabla M^*)(\hat{B}\cdot \nabla M)-i\frac{b_{5,1}}{2}|M|^2\left [M^*(\hat{B}\cdot \nabla M)-(\hat{B}\cdot \nabla M^*)M\right ]\nonumber
\\
&+&\frac{ib_{5,3}}{2}\left[(\nabla^2M^*) \hat{B}\cdot \nabla M- \hat{B}\cdot \nabla M^*(\nabla^2 M)\right]+a_{6,0}|M|^6+a^{(0)}_{6,2}|M|^2|\nabla M|^2\nonumber
\\
&+&a^{(1)}_{6,2}|M|^2(\hat{B}\cdot \nabla M^*)(\hat{B}\cdot \nabla M)+a_{6,4}|\nabla^2 M|^2 +... 
\end{eqnarray}

The magnetic field produces two distinguishable effects on the GL expansion. First, it allows terms even in $\hat{B}$ that are responsible for the explicit separation of transverse and parallel derivatives, as it is expected to occur in any theory where the rotational symmetry is broken by an external vector. These are the terms with coefficients $a^{(1)}_{i,j}$, which have similar structures to those with coefficients $a^{(0)}_{i,j}$,  except that the gradient operator is replaced by the projection of the gradient along the field. 
Second, the symmetries of the theory also allow to construct B-dependent terms that are linear in $\hat{B}$. These are the structures with coefficients $b_{i,j}$. As $B$ is odd under the T symmetry, the rest of the structure has to be also odd under T, hence odd in the pseudoscalar order parameter.  Even though these terms are permitted from general symmetry arguments, they are not a common feature of theories with an external vector, but they exist instead when the system exhibits nontrivial topology. We shall see below that, as it was showed in \cite{Incera}, in the MDCDW case, the existence of non-zero $b_{i,j}$ can be indeed traced back to the nontrivial topology manifested through the spectral asymmetry of the LLL fermions.

We call the readers attention to the fact that the anisotropy between transverse and parallel (to the magnetic field direction) vectors created by the explicit breaking of the rotational symmetry by the magnetic field in the MDCDW system is fundamentally different from the one created by the direction of the modulation in the DCDW case, where it is a result of spontaneous breaking of the rotational symmetry. This difference leads to quite different low-energy theories of the fluctuations in these two models.

Considering that the preferred density wave in the MDCDW case is a single-modulated density wave with its modulation vector parallel to the magnetic field, $M(z)=m e^{iqz}$,  $m \equiv -2G \Delta$, the free energy (\ref{GL-Free Energy}) can be written as
\begin{eqnarray}\label{GL-Free Energy-qdelta}
\mathcal{F}&=&a_{2,0}m^2+b_{3,1}qm^2+a_{4,0} m^4+a_{4,2}q^2m^2+b_{5,1}qm^4\nonumber
\\
&+&b_{5,3}q^3m^2+a_{6,0}m^6+a_{6,2}q^2m^4+a_{6,4}q^4m^2,
\end{eqnarray}
where $a_{4,2}=a^{(0)}_{4,2}+a^{(1)}_{4,2}$, $a_{6,2}=a^{(0)}_{6,2}+a^{(1)}_{6,2}$. In (\ref{GL-Free Energy-qdelta}), we keep up to sixth order terms to ensure the stability of the MDCDW phase in the mean-field approximation. 

It is important to point out that, straight derivations \cite{Gyory-Incera} show that the $a$ coefficients get contributions from all Landau levels $l$, while the $b$-coefficients do not get contributions from the higher Landau levels (hLL) $l>1$. 
This follows from the fact that the $b$-terms in (\ref{GL-Free Energy-qdelta}) are odd in q, which leaves the LLL modes as the only possible source of the $b$-terms. Indeed, the LLL contribution is not invariant under $q\to-q$, due to the asymmetry of the LLL modes (\ref{LLLspectrum}). In principle, the LLL part of the thermodynamic potential can have q-odd and q-even terms. Obviously, the $b$-terms come from the odd part. Such an odd-part is topological in nature, a fact that manifests in the existence of several anomalous quantities, like the anomalous part of the quark number, which is proportional to a topological invariant \cite{Tatsumi-2}, or the anomalous electric charge and the anomalous Hall current  \cite{Ferrer-PLB, Ferrer-NPB}, all of which are odd in $q$. 

In summary, the additional $a$ and $b$ terms have quite different origins. $a^{(1)}$-type terms will always appear in the presence of an external magnetic field, because they reflect the explicit breaking of the rotational symmetry produced by the field direction. On the other hand, the $b$ terms are associated to the topology of the modified fermion spectrum in the presence of the field.  As the LLL part of the thermodynamic potential is linear in the magnetic field $B$, so will be the $b$-coefficients.
 
The stationary equations from where the ground state solutions for $m$ and $q$ can be found are 
\begin{eqnarray}\label{SC-Delta}
\partial \mathcal{F}/\partial m&=&2m \{ a_{2,0}+2a_{4,0} m^2+3a_{6,0}m^4+q^2[a_{4,2}+2a_{6,2}m^2+a_{6,4}q^2]\nonumber
\\
&+& [b_{3,1}+2b_{5,1} m^2+b_{5,3}q^2]\}=0,
\end{eqnarray}

\begin{equation}\label{SC-q}
\partial\mathcal{F}/\partial q=m^2  \{2q[a_{4.2}+a_{6,2}m^2+2a_{6.4}q^2] +b_{3,1}+b_{5,1} m^2+3b_{5,3}q^2\}=0
\end{equation}

The minimum equations of the DCDW phase can be readily found from the zero-magnetic-field limit of (\ref{SC-Delta}), (\ref{SC-q}), where the $a^{(1)}_{i,j}$ and $b_{i,j}$ coefficients vanish.

Following \cite{Incera}, we now explore the theory beyond the mean-field approximation to check if the Landau-Peierls instability found in the absence of a magnetic field (the DCDW phase)  \cite{Tatsumi} is present here too. With this goal in mind, we investigate the low-energy thermal fluctuations that may affect the long-range order of the inhomogeneous ground state. Notice that in principle there can be fluctuations of the condensate magnitude and of the condensate phase, but we only need to care about fluctuations associated to the spontaneous breaking of global symmetries, as those are the ones that could in principle have soft modes that lead to the instability. In other words, to probe the instability of the ground state at arbitrarily low temperatures, the relevant fluctuations are those that can be excited at very low energies, i.e., those generated by the Goldstone bosons of the system. Hence, in our analysis, we do not consider the magnitude fluctuations because they are not associated to a Goldstone mode.

The symmetry group of the MDCDW phase is  $U_V(1)\times SO(2) \times R^2$, since the ground state of this phase spontaneously breaks the chiral symmetry $U_A(1)$ and the translation along z. Hence, there are two Goldstone bosons: the neutral pion, $\tau$, associated with the breaking of the chiral symmetry and the phonon, $\xi$, associated to the breaking of the translation symmetry. Now, the effect of the global transformations of these broken groups on the order parameter is 
\begin{equation}\label{SB-Pattern-Finite}
M(x) \rightarrow e^{i\tau}M(z+\xi)=e^{i(\tau+q\xi)}M(z),
 \end{equation}
from where one clearly sees that there is a locking between the chiral rotation and the z-translation. Therefore, we can always express them as two orthogonal combinations, one that leaves the order parameter invariant and one that changes it. As a consequence, there is only one legitimate Goldstone field in the MDCDW theory. One can arbitrarily choose it as either the pion, the phonon, or a linear combination of them. Henceforth, without loss of generality, we consider it to be the phonon. 

Let us consider now a small phonon fluctuation $u(x)$ on the order parameter and expand it about the condensate solution up to quadratic order in the fluctuation, 
\begin{equation}\label{Fluctuation}
M(x)=M(z+u(x))\simeq M_0(z)+M'_0(z)u(x)+\frac{1}{2} M''_0(z) u^2(x),
 \end{equation}
where $M_0(z)=\bar{m}e^{i\bar{q}z}$ is the ground state solution with $\bar{m}$ and $\bar{q}$ given as the solutions of (\ref{SC-Delta})-(\ref{SC-q}). 

Substituting (\ref{Fluctuation}) into (\ref{GL-Free Energy}), and keeping terms up to quadratic order in $u(x)$, we arrive at the phonon free energy 
\begin{equation}\label{Fluctuation-Free Energy}
\mathcal{F}[M(x)]=\mathcal{F}_0+v^2_z(\partial_z\theta)^2+v_\bot^2(\partial_\bot \theta)^2+\zeta^2(\partial_z^2\theta+\partial_\bot^2\theta)^2,
\end{equation} 
For convenience we wrote (\ref{Fluctuation-Free Energy}) in terms of the pseudo scalar $\theta=qmu(x)$, which is proportional to the phonon but with the dimension of a spin-zero field. Here
$\mathcal{F}_0=\mathcal{F}(M_0)$, $(\partial_\bot \theta)^2=(\partial_x \theta)^2+(\partial_y\theta)^2$ and $\zeta^2=a_{6.4}$. Notice that in deriving (\ref{Fluctuation-Free Energy}), the term linear in $\partial_z \theta$ cancels out after using (\ref{SC-q}).

The coefficients $v^2_z$, $v_\bot^2$ in (\ref{Fluctuation-Free Energy}) are given by
\begin{equation}\label{Parallel-Coef2}
v^2_z= a_{4.2}+\bar{m}^2 a_{6.2}+6\bar{q}^2a_{6.4}+3\bar{q}b_{5,3}
\end{equation}
\begin{equation}\label{Transverse-Coef2}
v_\bot^2= a_{4.2}+\bar{m}^2 a_{6.2}+2\bar{q}^2 a_{6.4}+\bar{q}b_{5,3}-a^{(1)}_{4.2}-\bar{m}^2a^{(1)}_{6.2}
\end{equation}  
They represent the squares of the parallel and transverse group velocities respectively. 

The fluctuation low-energy Lagrangian density is then
\begin{equation}\label{phononLagrangian}
\mathcal{L}_{\theta}=\frac{1}{2}[(\partial_0\theta)^2-v^2_z(\partial_z\theta)^2-v_\bot^2(\partial_\bot \theta)^2-\zeta^2(\partial_z^2\theta+\partial_\bot^2\theta)^2],
\end{equation}
from which we find the spectrum  
\begin{equation}\label{spectrum}
E\simeq\sqrt{v^2_zk^2_z +v_\bot^2k_\bot^2},
\end{equation}
with $k_\bot^2=k_x^2+k_y^2$.

The spectrum of the fluctuations is anisotropic and linear in both the longitudinal and transverse directions. It is easy to see that $v_z\neq 0$ because $a_{6.4}$ cannot be zero for the minimum solution to exist \cite{Nickel-2}. As for $v_\bot^2$, one can gather from (\ref{SC-q}) and (\ref{Transverse-Coef2}) that the $a^{(1)}_{i,j}$ and $b_{i,j}$ coefficients entering in the transverse group velocity serve to avoid the softness in the transverse direction normally seen in single-modulated phases like the DCDW.  Let us recall that in the DCDW phase there is no magnetic field and thus these coefficients are zero. In such a case, the remaining combination in (\ref{Transverse-Coef2}) vanishes due to the stationary condition (\ref{SC-q}), thereby leading to $v_\bot=0$. On the other hand, the lack of soft modes ensured by the additional coefficients in the MDCDW phase have remarkable consequences for the stability of the condensate, as will be shown below.  

In order to investigate the stability of the condensate against the fluctuations we need to calculate its average
\begin{equation}\label{averageM}
\langle M \rangle=\bar{m} e^{i\bar{q}z}\langle \cos \bar{q}u \rangle,
\end{equation}
with the average defined as 
 \begin{equation}\label{average}
\langle  ... \rangle =\frac{\int \mathcal{D}u(x) ... e^{-S(u^2)}}{\int \mathcal{D}u(x) e^{-S(u^2)}}
\end{equation}
where
 \begin{equation}\label{phonon action}
S(u^2)= T\sum_n \int^{\infty}_{-\infty} \frac{d^3k}{(2\pi)^3} [\omega^2_n+(v^2_zk^2_z +v_\bot^2k_\bot^2+\zeta^2 k^4)]\bar{q}^2\bar{m}^2u^2.
\end{equation}
denotes the finite-temperature effective action of the phonon and $\omega_n=2n\pi T$ the Matsubara frequency. 

Considering the relation
\begin{equation}\label{cos-exp-relation}
\langle \cos \bar{q}u \rangle =e^{-\langle (\bar{q}u)^2 \rangle/2}
\end{equation}
and using (\ref{average}), we find the mean square of the fluctuation as
\begin{eqnarray}\label{Fluctuation-2}
\langle \bar{q}^2u^2 \rangle &=&\frac{1}{(2\pi)^2}\int_{0}^\infty dk_\bot k_\bot \int_{-\infty}^\infty dk_z \frac{T}{\bar{m}^2(v^2_zk^2_z +v_\bot^2k_\bot^2+\zeta^2 k^4)}\nonumber
\\
&\simeq& \frac{\pi T}{\bar{m}\sqrt{v^2_zv_\bot^2}}.
\end{eqnarray}
where we took into account that the lowest Matsubara mode is dominant in the infrared. 

From (\ref{Fluctuation-2}), (\ref{cos-exp-relation}) and (\ref{averageM}) we can see that $\langle M \rangle \neq 0$ since $\langle \bar{q}^2u^2 \rangle$ is finite. This implies that the MDCDW system does not exhibit the Landau-Peierls instability, meaning that at $B \neq 0$ the fluctuations do not wipe out the condensate at arbitrarily low $T$. As can be gathered from our derivations, the lack of Landau-Peierls instabilities in the presence of a magnetic field is a direct consequence of the stiffening of the spectrum in the transverse direction, which in turn is due to the explicit breaking of the rotational symmetry by the external field. 

We should point out that the lack of Landau-Peierls instabilities in the presence of a magnetic field will not be changed by a nonzero current quark mass, since this property comes from the effect of the magnetic field on the low energy behavior of the phonon, which remains a Goldstone boson even at nonzero quark masses.

%%%%%%%%%%%%%%%%%%%%%%%%%%%%%%%%%%%%%%%%%%
\section{ Hybrid Propagation Modes in the MDCDW Medium}

In this section we investigate the propagation of electromagnetic waves in the MDCDW phase by going beyond the mean-field approximation to study the effects of the phonon fluctuations when the MDCD medium interacts with photons. This question is not only of fundamental interest to understand the properties of matter-light interaction in the MDCDW medium, but it may be also relevant to explain the stability of NS in very active $\gamma$-ray regions, as will be discussed in Section 6.

In the previous section we saw that the low-energy theory of the fluctuations in the MDCDW phase is given by (\ref{phononLagrangian}). This result considered a background magnetic field interacting with the quark medium, but assumed no other electromagnetic field was present. However, there are situations where the MDCDW medium may be penetrated by photons and we need to understand if their interaction with the medium can produce new physical effects.

When photons are present in the MDCDW medium, the low-energy theory of the fluctuations acquires the following additional contributions
\begin{equation}\label{extraEMterms}
\mathcal{L}_{A-\theta}=-\frac{1}{4}F_{\mu\nu}F^{\mu \nu}+ J^\mu A_\mu+\frac{\kappa}{8} \theta_0(x) F_{\mu\nu}\tilde{F}^{\mu \nu}+\frac{\kappa}{8} \theta(x) F_{\mu\nu}\tilde{F}^{\mu \nu},
\end{equation}
The first two terms are the conventional Maxwell and ordinary 4-current contributions respectively, the latter obtained after integrating out the fermions in the original MDCDW effective action \cite{Ferrer-PLB, Ferrer-NPB}. The last two terms are the axial anomaly with background axion field $\theta_0(x)=mqz$ and its (phonon-induced) fluctuation $\theta(x)$. Here $\kappa=2\alpha/\pi m$. 

The combined Lagrangian $\mathcal{L}=\mathcal{L}_{\theta}+\mathcal{L}_{A-\theta}$ effectively describes the low-energy theory of an axion field $\theta(x)$ interacting nonlinearly with the photon via the chiral anomaly. Let us now assume that a linearly polarized electromagnetic wave, with its electric field $\mathbf{E}$ parallel to the background magnetic field $\mathbf{B}_0$, propagates in the MDCDW medium \cite{Polariton}. The field equations of this theory are:
\begin{eqnarray} \label{CoupleEq}
&\mathbf{\nabla} \cdot \mathbf{E}=J^0+\frac{\kappa}{2}\nabla \theta_0 \cdot \mathbf{B}+\frac{\kappa}{2}\nabla \theta \cdot \mathbf{B}, \label{1}
 \\
&\nabla \times \mathbf{B}-\partial \mathbf{E}/\partial t=\mathbf{J}-\frac{\kappa}{2} (\frac{\partial \theta}{\partial t} \mathbf{B}+\nabla \theta\times \mathbf{E}),  \label{2}
\\
&\mathbf{\nabla} \cdot \mathbf{B}=0, \quad \nabla \times \mathbf{E}+\partial \mathbf{B}/\partial t=0 \label{3}
\\
&\partial_0^2 \theta - v_z^2 \partial_z^2\theta-v_\bot ^2 \partial_\bot^2\theta+\frac{\kappa}{2} \mathbf{B} \cdot \mathbf{E}= 0,
\end{eqnarray} 
which contains terms coupling the axion with the photon. In (\ref{CoupleEq}), $\mathbf{B}$ is the total magnetic field, meaning the background field plus the wave magnetic field.

Since we are interested in applications to NS, we should consider a neutral medium, hence we assume that $J^0$  contains an electron background charge that ensures overall neutrality
\begin{equation}\label{EMterms}
J^0+\frac{\kappa}{2}\nabla \theta_0 \cdot \mathbf{B}+\frac{\kappa}{2}\nabla \theta \cdot \mathbf{B}=0.
\end{equation}

 The linearized field equations can then be written as
\begin{eqnarray} \label{Wave-Eq}
&\partial^2\mathbf{E}/\partial t^2=\mathbf{\nabla}^2 \mathbf{E}+\frac{\kappa}{2} (\partial^2\theta/\partial t^2)\mathbf{B}_0
\\
&\partial^2\theta/\partial t^2- v_z^2 (\partial^2\theta/\partial z^2)-v_\bot ^2 (\partial^2\theta/\partial x^2+\partial^2\theta/\partial y^2)+\frac{\kappa}{2} \mathbf{B}_0 \cdot \mathbf{E}= 0.
\end{eqnarray}

Their solutions describe two hybridized propagating modes of coupled axion and photon fields that we call axion polaritons (AP), borrowing the term from condensed matter. In general, polaritons are hybridized propagating modes that emerge when a collective mode like phonons, magnons, etc., couples linearly to light.

The energy spectrum of the hybrid modes are 
\begin{equation}\label{Frequencies-1}
\omega^2_{0}=A-B, 
\end{equation}

\begin{equation}\label{Frequencies-2}
\quad \omega^2_{m}=A+B
\end{equation}

with
\begin{equation}\label{A}
A=\frac{1}{2}[p^2+q^2+(\frac{\kappa}{2} B_0)^2],
\end{equation}
\begin{equation}\label{B}
B=\frac{1}{2}\sqrt{ [p^2+q^2+(\frac{\kappa}{2} B_0)^2]^2-4p^2q^2},
\end{equation}
and $q^2=v_z^2 p_z^2+v_\bot^2 p_\bot^2$. 

From (\ref{Frequencies-1})-(\ref{B}) we identify $\omega_{0}$ as the gapless mode and $\omega_{m}$ as the gapped mode with field-dependent gap 
\begin{equation}\label{AP-Mass}
\omega_{m}(\vec{p}\rightarrow0)=m_{AP}=\alpha B_0/\pi m
\end{equation}

Similarly coupled modes of axion and photon have been found in topological magnetic insulators \cite{Polaritons-CM}, underlining once again the striking similarities between MDCDW quark matter and topological materials in condensed matter.

%%%%%%%%%%%%%%%%%%%%%%%%%%%%%%%%%%%%%%%%%%
\section{Axion Polariton and the Missing Pulsar Problem}

The fact that the MDCDW medium can create massive AP's when it is bombarded with electromagnetic radiation may have important implications for the physics of NS in the galactic center  (GC) \cite{Polariton}. A long-standing puzzle in astrophysics, known as the missing pulsar problem, refers to the failed expectation to observe a large number of pulsars within 10 pc of the galaxy center. Theoretical predictions have indicated that there should be more than $10^3$ active radio pulsars in that region \cite{G-M-P}, but these numbers have not been observed. This paradox has been magnified by pulse observations of the magnetar SGR J1745-2900 detected by the NuSTAR and Swift satellites \cite{Mori}-\cite{Magnetar-Detect}. These observations revealed that the failures to detect ordinary pulsars at low frequencies cannot be simply due to strong interstellar scattering, but instead should be connected to an intrinsic deficit produced by other causes. 

Furthermore, as pointed out in \cite{Dexter}, the detection of the young ($T \sim 10^4$ yr) magnetar SGR J1745-2900 indicates high efficiency for magnetars formation from massive stars in the GC, because it will be unlikely to see a magnetar unless magnetar formation is efficient there. In fact, it has been argued that the detection of SGR 1745-2900, with a projected offset of only $0.12$ pc from the GC, should not have been expected unless magnetar formation is efficient in the GC with an order unity efficiency \cite{Dexter}, and that the missing pulsar problem could be explained as a consequence of a tendency to create short-lived magnetars rather than long-lived ordinary pulsars. On the other hand, there is evidence that several magnetars are associated with massive stellar progenitors ($M>40 M_\odot$) \cite{Figer}, a fact that supports the idea that magnetars formed in the GC could be very massive compact objects made of quark matter. These massive magnetars can be 2$M_\odot$ quark stars with inner magnetic field $B=10^{17}$ G. Although the original argument for the existence of quark stars was based on the stability of strange quark matter, in recent years it was demonstrated \cite{Holdom}, using a phenomenological quark-meson model that includes the flavor-dependent feedback of the quark gas on the QCD vacuum, that $u$-$d$ quark mater is in general more stable than strange quark matter, and it can be more stable than the ordinary nuclear matter when the baryon number is sufficiently large. Based on this result, for the analysis below, we shall consider the hypothesis that the massive magnetars in the GC are two-flavor quark stars in the MDCDW phase. 

The Milky Way GC is a very active astrophysical environment with numerous $\gamma$-ray emitting point sources \cite{GR}. Extragalactic sources of GRB show an isotropic distribution over the whole sky flashing with a rate of 1000/year. The energy output of these events is $\sim 10^{56}$ MeV, with photon energies of order $ 0.1 - 1$ MeV \cite{GRB}, meaning that each one of these events can produce $10^{56}$ or more photons. If we assume that only $10 \%$ of these photons reach the star, which is a conservative estimate if the star is in the narrow cone of a GRB beam, about $10^{55}$ of those photons can reach the NS. 

For fields $B=10^{17}$ G, the mass gap $m_{AP}$ of the gapped AP is in the range $[0.06, 0.3]$ MeV for corresponding parameter intervals $\mu \in [340.1, 342.5]$ MeV and $m \in [23.5, 4.7]$ MeV \cite{Gyory-Incera}. Hence, one can gather that many of the photons reaching the interior of a quark star in the MDCDW phase can have enough energy to propagate inside as gapped AP's. The conversion of a large number of $\gamma$-photons into AP's once they hit the NS interior can take place through the so-called Primakoff effect \cite{Primakoff}. The Primakoff effect is a mechanism that can occur in theories that contain a vertex between a scalar or a pseudoscalar and two photons, so that via this vertex and in the presence of background electric or magnetic fields, the photon can be transformed into these bosons. In the context of the MDCDW dense quark matter, the Primakoff effect allows the incident photons to be transformed into AP's thanks to the anomalous axion-two-photons vertex and the existence of a background magnetic field. This effect can produce a large number of gapped AP's, which being bosons, will be gravitationally attracted to the center of the star where they will accumulate with high density.

If the number of AP's that accumulates in the star's center is higher than the 
Chandrasekhar limit \textit{for these bosons}, the AP's will create a mini black hole in the star center that will destroy the host NS, leaving a remnant black hole. We explored this possibility in \cite{Incera}, where we considered the Chandrasekhar limit that determines the number of AP's required to induce the collapse, ignoring the gravitational energy associated with the quarks. For boson particles, this limit is given by \cite{Ch-Bosons-1, Ch-Bosons-2}
\begin{equation}\label{N-Cha}
N_{AP}^{Ch}= \left(\frac{M_{pl}}{m_{AP}}\right)^2=1.5 \times 10^{44} \left(\frac{MeV}{m_{AP}}\right)^2
\end{equation}
where $M_{pl}=1.22\times 10^{19}$ GeV is the Planck scale. Using the largest AP mass $m_{AP}=0.3$ MeV for the $B=10^{17}$ G field, we find $N_{AP}^{Ch}=1.7\times10^{45}$. This implies that if just $10^{-8} \%$ of the $10^{55}$ photons reach the star with energies $\sim 0.3$ MeV or larger, they can in principle generate a large enough number of AP's to produce a mini black hole in the star's center and induce its collapse.  Similarly, for an AP mass $m_{AP}=0.06$ MeV, we find $N_{AP}^{Ch}=4.2\times10^{46}$, so in this case $10^{-7} \%$ of the total number of photons will have to reach the star to create the conditions for the collapse. Notice that this mechanism is purely a bosonic effect, since it is related to the Chandrasekhar limit of the bosons. 

We should point out that the likelihood of reaching the Chandrasekhar limit in the star interior is not just determined by the number and energies of the $\gamma$-rays hitting the star, but also by the capacity of these photons to penetrate the quark medium and then generate a large enough number of AP's that get trapped by the star gravity.  Hence, for the above AP mechanism to be operative, one has to estimate the $\gamma$-rays attenuation in the MCDCW quark medium and use it to determine whether the star can trap or not the AP's that form in its interior \cite{Incera}. 

In a medium, $\gamma$-rays are mainly attenuated by their interaction with electrons. The main process driving the attenuation in a NS is Compton scattering. The attenuation at a given depth can be found from the formula
\begin{equation}\label{Attenuation-Formula}
 I=I_0 e^{-\sigma n_e L},
  \end{equation}
where $I_0$ is the incident radiation intensity, $I$ the intensity at a thickness L inside the medium, $\sigma$ the cross section of Compton scattering and $n_e$ the electron number density. 
In a quark star, to reach the quark medium the $\gamma$ rays have to cross an electron cloud of thickness a few hundreds $\textrm{fm}$, since quark stars exhibit a macroscopic quark-matter surface shrouded with this very thin electron cloud  \cite{Alcock}. The quark-star surface acts as a membrane that allows only ultrarelativistic matter to escape: photons, neutrinos, electron-positron pairs and magnetic fields. For the incoming $\gamma$ rays to reach the quark matter medium and activate the Primakoff effect, they first need to go through the electron cloud without big attenuation losses.

The formula that gives the Compton scattering cross section is known as the Klein-Nishina formula \cite{Klein-Nishina, Longair} and is given by
 \begin{eqnarray}
\sigma&=&\frac{3e^4}{48\pi \epsilon_0^2 m_e^2 c^4}  \left [\frac{1}{x}\left (1-\frac{2(x+1)}{x^2} \right ) ln (2x+1)+\frac{x+8}{2x^2}-\frac{1}{2x(2x+2)^2} \right ]
\\
&=&2.49 \times 10^{-25}\left [\left (\frac{x^2-2(x+1)}{x^3} \right ) ln (2x+1)+\frac{x+8}{2x^2}-\frac{1}{2x(2x+2)^2} \right ] cm^2
\nonumber
\label{K-N Formula}
\end{eqnarray}   
where $x=\omega / m_e c^2$ is the ratio between the photon energy and the rest energy of the electron. For the maximum incident photon energy $\omega \approx 1$ MeV, $x\approx2$ and the cross section is
\begin{equation}\label{cross-section-value}
\sigma\approx 2.58 \times 10^{-25} cm^2
  \end{equation}

The electron number density of the cloud can be found from \cite{Alcock} 
\begin{equation}\label{N-e}
n_e=\frac{9.49 \times 10^{35}cm^{-3}}{\left [ 1.2 \left (z/10^{-11} cm \right ) +4 \right ]^3} 
  \end{equation}
Here $z$ is the heigh above the quark surface.

From Eqs. (\ref{Attenuation-Formula}), (\ref{cross-section-value}) and (\ref{N-e}) we obtain that the ratio of intensities for $L\approx z  \approx 300 fm$ is $I/I_0\approx 0.983$, which shows that for $1$MeV $\gamma$ rays the attenuation is negligible. A similar calculation for the least energetic incident $\gamma$-ray, with $0.1$ MeV, still shows small attenuation $I/I_0\approx 0.64$. In the case of hybrid stars the situation is different since the $\gamma$ rays have to cross several kilometers of hadronic matter with a relatively large electron density before reaching the quarks in the core. It can be proved that in this case the $\gamma$ radiation will be absorbed in a distance of less than a hundred of $fm$ into the mantle.

Back to the quark star case, once the $\gamma$ rays reach the quark medium they are converted to AP via the Primakoff effect and a natural question immediately follows:  Are these AP trapped inside the star? To answer this question we need to compare the velocity of the AP with the star's escape velocity $v_e/c=\sqrt{2GM_{star}/c^2R_{star}}$. For a star with $M_{star}=2 M_\odot$ and $R_{star}=10$ km, we have $v_e= 0.8 c$. The velocity $v_{AP}$ that an AP of mass $m_{AP}$ can reach depends on the energy $E$ it acquires from the  incident $\gamma$-rays
\begin{equation}\label{Axion-Max-Vel}
v_{AP}/c=\sqrt{1-\left( \frac{m_{AP}c^2} {E} \right)^2}. 
\end{equation}
For instance, for $m_{AP}c^2=0.3$ MeV, all the AP with energy $E < 0.5$ MeV cannot escape. Similarly, if $m_{AP}c^2=0.06$ MeV, the AP's with energies $E < 0.1$ MeV will be gravitationally trapped. This implies that for incident $\gamma$-photons in the energy interval $(0.1,0.5)$ MeV, there will always be AP's that will be trapped. The use of 2$M_\odot$ stars in the escape velocity is motivated by recent indications \cite{Nature2020} that the heaviest neutron stars, with masses $\sim 2 M_\odot$, should have deconfined quark-matter inside. 

The constraint in the interval of photon energies needed to generate APs that will be trapped in turn affects the estimate of the percentages of incident photons needed to reach the Chandrasekhar limit. If we conservatively assume that the number of photons per energy is the same throughout the entire interval of $\gamma$-ray energies, then we can estimate that photons in the energy interval $(0.3,0.5)$ MeV roughly represent $22\%$ of the total number of incident photons that reach the quark medium, i.e., about $\approx 10^{54}$ photons. Of these photons, only $10^{-7} \%$, are needed to generate enough AP's to reach the Chandrasekhar limit and induce the star's collapse. We then conclude that the AP mechanism to collapse the star by creating a mini-black hole from the creation and subsequent accumulation of AP particles in the star's center is viable for quark stars and can serve to explain the missing pulsar problem. 

There are several reasons why the presence of a magnetic field is crucial for the AP mechanism to work. First, because a background magnetic field is needed for the density wave phase of quarks to be stable against low-energy fluctuations \cite{Incera}, second, because a background magnetic field is needed to create APs through the Primakoff effect \cite{Polariton}, and third, because the AP gap is proportional to the magnetic field \cite{Polariton}. It is worth to mention that the AP mechanism does not require unrealistically large magnetic fields to be viable. Fields of magnitude $10^{16}-10^{17}$ G are enough to make the MDCDW phase energetically favored over the chirally restored one at intermediate densities. These are plausible fields for the interior of magnetars, whose surface magnetic fields can be as high as $10^{15}$G. All these facts, together with the intense $\gamma$-ray activity in the galaxy center, create the conditions needed for the collapse of those short-lived magnetars via the AP scenario.

%%%%%%%%%%%%%%%%%%%%%%%%%%%%%%%%%%%%%%%%%%
\section{MDCDW condensate versus Magnetically Catalyzed Chiral Condensate}

It is well known that in a system of massless charged fermions in a magnetic field, the dimensional reduction in the infrared dynamics of the particles in the LLL favors the formation of an homogeneous particle-antiparticle chiral condensate even at the weakest attractive interaction between fermions. This phenomenon is due to the fact that there is no energy gap between the infrared fermions in the LLL and their antiparticles in the Dirac sea. This phenomenon is known as the magnetic catalysis of chiral symmetry breaking (MC$\chi$SB) \cite{MC-1}- \cite{MC-6}. The MC$\chi$SB is a universal phenomenon that has been tested in many different contexts \cite{MC-7}- \cite{MC-13}. 

In the original studies of the MC$\chi$SB \cite{MC-1}-\cite{MC-13}, the catalyzed chiral condensate was assumed to generate only a
fermion dynamical mass. However, it was later shown that in QED \cite{ferrerincera-1, ferrerincera-2}, the MC$\chi$SB inevitably leads also to the emergence of a dynamical anomalous magnetic moment (AMM). The reason is that the AMM does not break any symmetry that has not already been broken by the chiral condensate. The dynamical AMM in massless QED leads, in turn, to a non-perturbative Lande g-factor and a Bohr magneton proportional to the inverse of the dynamical mass. The induction of the AMM also yields a non-perturbative Zeeman effect \cite{ferrerincera-1, ferrerincera-2}. 

Just as in QED, the magnetically catalyzed ground state of a NJL model of massless quarks at subcritical coupling turns out to be actually richer than previously thought with the emergence of two homogeneous condensates, the usual $\langle \overline{\psi}\psi\rangle$ and a magnetic moment condensate $\langle \overline{\psi}\Sigma^3\psi\rangle$ aligned with the magnetic field direction \cite{AMM-NJL}. An effect of the magnetic moment is to significantly enhance the critical temperature for chiral symmetry restoration. 

The above examples assumed zero chemical potential. A chemical potential can affect the picture significantly because once the density becomes different from zero and a Fermi surface is formed, the energy cost to pair particles with antiparticles grows, so that eventually, the pairing is not energetically favored any longer. At nonzero chemical potential, the MC$\chi$SB then occurs until $\mu$ reaches a critical value at which a first-order phase transition takes place and the chiral symmetry is restored \cite{Leung-mu}.

In Section 2, we saw that the MDCDW condensate can be formed even in the subcritical coupling regime, as long as the chemical potential is nonzero. That means that there is a region of chemical potentials where the MDCDW chiral condensate and the homogeneous MC$\chi$SB condensate compete with each other. Which of them is more energetically favored can be gathered from Fig. 3, where the plots of the free energy vs the chemical potential are displayed for the MC$\chi$SB phase (yellow lines) and the MDCDW phase (blue lines), at different magnetic fields and/or couplings. Clearly, the spatially inhomogeneous condensate wins over the homogeneous one in the entire region of chemical potentials in all the situations. 

\begin{figure}
\begin{center}
\begin{tabular}{ccc}
  \includegraphics[width=4.2cm]{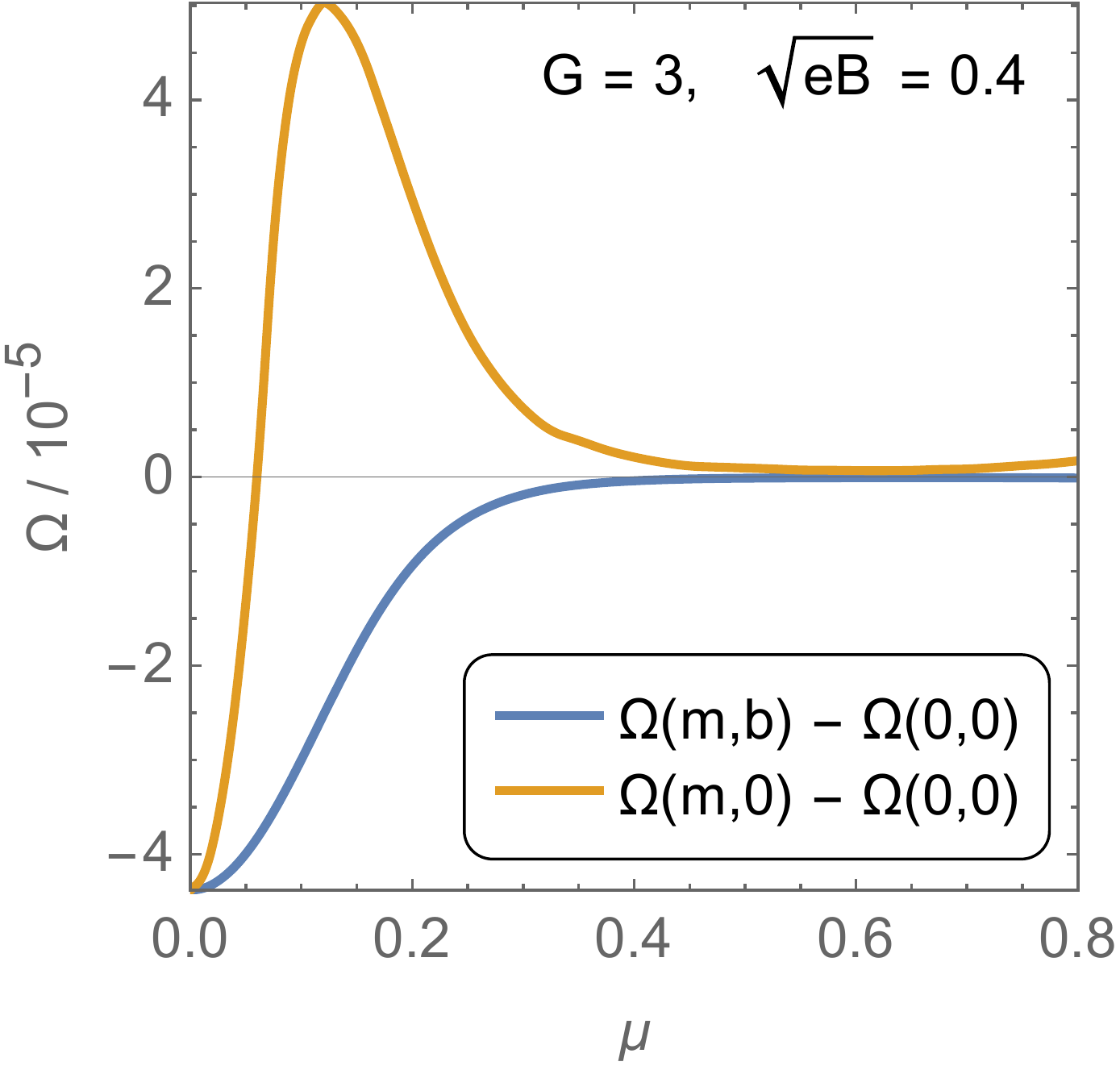} & \includegraphics[width=4.2cm]{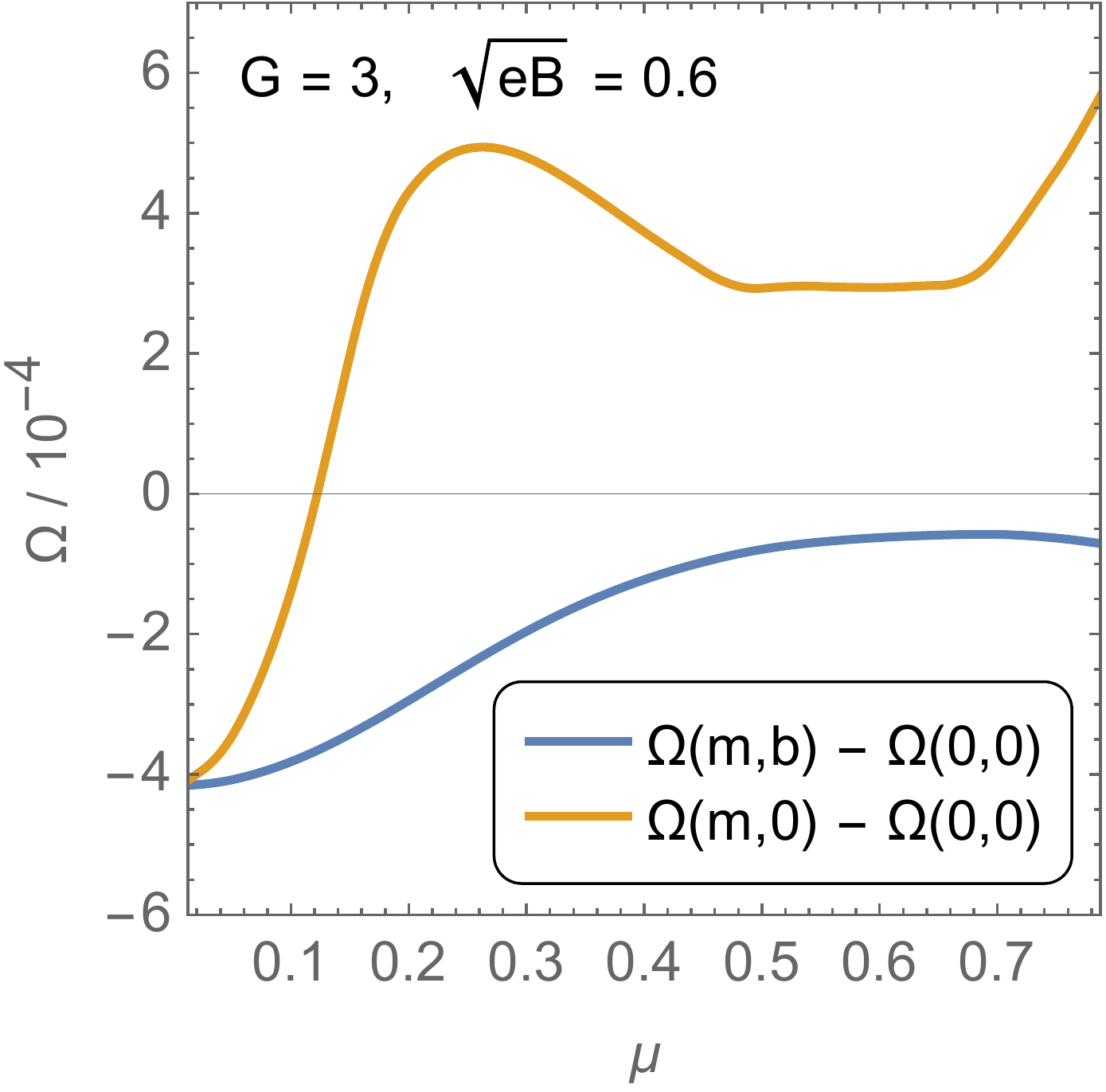} & \includegraphics[width=4.2cm]{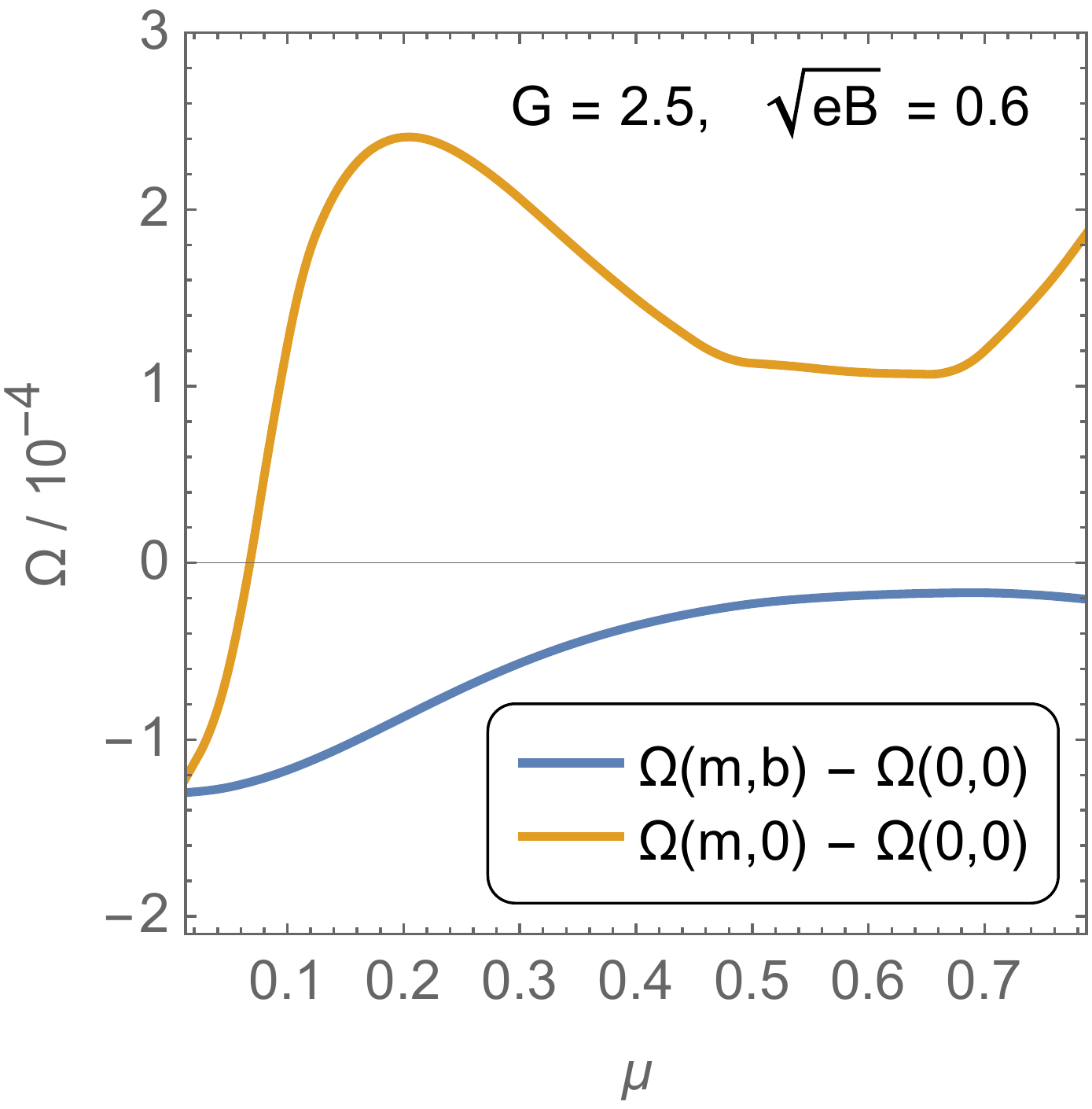} \\
(a) & (b) & (c) \\
  \\
  \end{tabular}
    \end{center}
    \caption{Comparison of free energies for two phases at sub-critical couplings: a homogeneous phase with a magnetically catalyzed condensate (yellow line) and a spatially inhomogeneous phase (blue line) with an MDCDW condensate.}
     \label{Fig-3}
\end{figure}

Comparing Fig. 3 (a) and (b) plots, one can see that a larger magnetic field decreases the free energy of the inhomogeneous phase and increases that of the homogeneous one. A similar behavior occurs at fixed field $B$ but different couplings, as can be seen from (b) and (c). Here, the separation between the two free energies increases with the coupling, clearly favoring the MDCDW phase. These results underline how robust the MDCDW is even at sub-critical coupling, an effect that can be connected to the topological contribution to the free energy from the LLL quarks.

It is worth to stress another difference between these two phases. While a driven factor in the MC$\chi$SB case is the LLL infrared dynamics, as already pointed out, in the MDCDW phase there is a connection between ultraviolet (UV) and infrared (IR) phenomena. The appearance of $\Omega_{anom}$ (\ref{Omega-anom}) in the thermodynamic potential is a consequence of the regularization of the high-energy modes in the difference of two ill-defined sums from which the anomalous and the finite medium contributions are extracted \cite{Klimenko}. Since the anomalous term contributes to the gap equation for $q$, whose origin is IR because it comes from the quark-hole pairing, we have that the UV physics and the IR properties of the system are interrelated.

Finally, we should comment on the fact that while in the original MC$\chi$SB phenomenon the condensate increases with the magnetic field, more recently it was found that if the effect of the magnetic field on the coupling constant is taken into consideration, the chiral condensate actually decreases with the magnetic field, a phenomenon known as inverse magnetic catalysis \cite{IMC-1}-\cite{ IMC-5}. It remains as an open and interesting question what will be the consequences of including the effect of the magnetic field on the strong coupling constant for the inhomogeneous condensate of the MDCDW phase.

%%%%%%%%%%%%%%%%%%%%%%%%%%%%%%%%%%%%%%%%%%
\section{Conclusions}

In this paper we have reviewed the main physical characteristics of the MDCDW phase of dense quark matter and its possible connection with the astrophysics of NS. One main attribute of this phase is its non-trivial topology, which is due to the combined effect of the density wave ground state and the dimensional reduction produced by the magnetic field on the LLL, which together give rise to an asymmetric spectrum for the LLL modes. As a consequence, the MDCDW phase displays anomalous properties like an anomalous electric charge that depends on the applied magnetic field and the modulation $q$, a non-dissipative anomalous Hall current, and magnetoelectricity. 

The topological nature of the the MDCDW phase is also reflected in the matter-light interactions and how they affect the propagation of photons in this medium which occurs via axion polaritons, a transport behavior that could help to explain the so-called missing pulsar problem in the GC.

A very important feature of the MDCDW phase is its stability against thermal phonon fluctuations at arbitrarily small temperatures. In other words, this system is protected against the Landau-Peierls instability \cite{Peirls, Landau} that usually erodes single-modulated phases in three spatial dimensions leading to the lack of a long-range order. The lack of the instability is due to magnetic field-induced terms in the low-energy GL expansion, some of which have a topological origin, since they are connected to the spectral asymmetry, and some of which are just the effect of the explicit breaking of the rotational symmetry by the magnetic field. The lack of Landau-Peierls instabilities in the MDCDW phase makes this phase particularly robust and hence a good candidate for the inner matter phase of neutron stars.

Although the emphasis of this paper has been in NS, the results of this review can also be of interest for Heavy Ion Collision (HIC) physics. Future  HIC experiments, plan to explore the region of lower temperatures and higher densities, and in doing that, they will certainly generate strong magnetic and electric fields in their off-central collisions, so opening a much more sensitive window to look into a very challenging region of QCD \cite{NICA}. For example,
the second phase of the RHIC energy scan (BES-II) \cite{Odyniec}, the planned experiments at the Facility for Antiproton and Ion Research (FAIR) \cite{1607.01487} at the GSI site in Germany, and the Nuclotron-based Ion Collider Facility (NICA) \cite{PRC85, Toneev} at JINR laboratory in Dubna, Russia, are all designed to run at unprecedented collision rates to provide high-precision measures of observables in the high baryon density and lower temperature region. 

Searching for signals of inhomogeneous quark phases in these planned experiments is a necessary step to probe their realization in this yet unexplored region of the QCD phase map. Recently, a proposal to detect those signatures has been discussed in \cite{PRL2021}. The idea is that in regimes with periodic spatial modulation, particles can have a "moat" spectrum, where the minimum of the energy is not at zero momentum, but lies over a sphere at nonzero spatial momentum. On the base that particle distribution with a moat spectrum should peak at nonzero momentum \cite{NPA1005}, the authors of \cite{PRL2021} argue that this feature can leave distinctive signatures in the production of particles and their correlations, which are measurable in heavy-ion collisions. The properties of the MDCDW phase discussed in this review, together with the upcoming findings of the range of critical temperatures at which the condensate evaporates \cite{Gyory-Incera}, can serve to guide the experiments to better pinpoint the region of parameters where the signatures of a moat spectrum are most likely to be detected. 

%can serve to inform those and other potential signatures in tfor the potential detection of  of It is then very relevant to carry out detailed theoretical investigations about the quark phases that can be realized at intermediate densities and finite temperatures. For the MDCDW phase, it is important to find the critical temperature at which this inhomogeneous condensate evaporates at each chemical potential, so to have a better idea of where in the QCD phase map the MDCDW phase will be localized. Second, it is also needed to investigate the characteristics of the first-order phase transition from the hadronic to the MDCDW phase produced by increasing the baryonic chemical potential. The role of the magnetic field in this transition can make a difference as discussed in \cite{Phase-Trans} due to the presence of the anomalous pressure, which is proportional to $\mu qB$, and hence a sufficiently high magnetic field can produce a critical value of the baryonic chemical potential significant different from the one driving the phase transition to a free quark gas. 

Finally, we should call the reader's attention to the fact that the anomalous effects of the MDCDW phase share many similarities with topological condensed matter systems as topological insulators \cite{Qie-PRB78}, where $\theta$ depends on the band structure of the insulator; Dirac semimetals \cite{Dirac-SM, Dirac-SM-2,  Dirac-SM-3, Dirac-SM-4}, a 3D bulk analogue of graphene with non-trivial topological structures; and WSM \cite{WeylSem}, where the derivative of the angle $\theta$ is related to the momentum separation between the Weyl nodes. Therefore, the discovery of new physical properties of these materials can shed light on the physics governing the challenging region of strongly coupled QCD, thereby inspiring new strategies to probe the presence of the MDCDW and other suitable phases in NS and HIC. 

%%%%%%%%%%%%%%%%%%%%%%%%%%%%%%%%%%%%%%%%%%
\acknowledgments{This work was supported in part by NSF grant PHY-2013222. We are grateful to W. Gyory for helpful discussions and for his help with several graphs.}

%=====================================
% References, variant A: internal bibliography
%=====================================
\end{document}